\documentclass[twocolumn,preprintnumbers,amsmath,amssymb]{revtex4}
\usepackage{epsfig}
\usepackage{graphicx}
\begin{document}\hbadness=10000

\title{Statistical Hadronization Probed by Resonances}
\author{Giorgio Torrieri and Johann Rafelski}
\affiliation{
Department of Physics, University of Arizona, Tucson, Arizona 85721, USA}
\date{May, 2003}
\begin{abstract}
\noindent
We study to what extent  a measurement of the $m_\bot$ spectra
 for hadrons and their resonances 
can resolve ambiguities in the statistical model description
of particle production. 
We describe in a quantitative analysis  how physical
assumptions about the freeze-out geometry and dynamics influence 
 the particle spectra.
Considering ratios of $m_\bot$ distribution of 
 resonance-particle  ratios (such as $K^*/K$, $\Sigma^*/\Lambda, \eta'/\eta$)
we observe significant  sensitivity
to fireball freeze-out geometry and flow dynamics.

PACS number(s): 12.38.Mh, 25.75.-q, 24.10.Pa
\end{abstract}
\vspace{+0.5cm}


\maketitle
\section{Introduction}
The Fermi statistical model of particle production~\cite{Fer50,Pom51,Lan53,Hag65}
 has been used extensively in the field of relativistic heavy ion collisions 
~\cite{jansbook}.
Particle abundancies and spectra both at Super Proton Synchrotron (SPS) 
~\cite{PBM99,NA57,becattini,van-leeuwen,ourspspaper,sqm2001,PBM01} and Relativistic Heavy Ion Collider (RHIC) 
~\cite{burward-hoy,castillo,PBMRHIC,bugaev_freeze,florkowski,rafelski2002} energies have been analyzed in this way.  
The quality of fits to experimental results  was
such that it became possible to discuss hadronization conditions 
quantitatively, but the conclusions of the groups differ.
For example  the values of temperature
range from as low as 110 MeV~\cite{burward-hoy,castillo,van-leeuwen,bugaev_freeze}
to 140 MeV~\cite{ourspspaper,rafelski2002}
to as high as 160, 170, and 180 MeV
\cite{florkowski,PBM99,PBM01,PBMRHIC,castillo,becattini} at both SPS and RHIC energies. 

These differences  are on a closer inspection not very surprising, since 
the (tacit) assumptions made about hadronization mechanisms  differ.
However, this means that before we can say that the freeze-out temperature has
been determined, we must understand precisely the origins
of these differences, and proceed to ascertain which model is applicable.
We shall suggest experimental observables which will be particularly
sensitive to the differences between the hadronization scenarios, in the hope
that further experimental study will allow to understand the statistical hadronization
mechanism. 

We begin with  an overview of the differences between 
hadronization scenarios and their relation to  the physical
assumptions used. Every model discussed here has been extensively
studied before, and has gained acceptance of some part of the heavy ion
community. The theoretical principles that we invoke are well 
understood, and the methods we use can be found  scattered in literature.
We shall concentrate here on an analysis of resonances
produced in a heavy ion collision.
Direct detection of hadronic short-lived resonances has become  possible through invariant
mass reconstruction~\cite{fachini,vanburen,markert,friese}.
 Resonances  have already been proposed as
candidates for differentiating between freeze-out models~\cite{resonances}.  
Resonances  are  a sensitive probe of the freeze-out temperature, since the  ratio
of yields  of particles with the same quark composition is 
 insensitive to both fugacities and phase space occupancies, and mass differences
are greater than the hadronization temperature considered.

Here 
we shall develop this reasoning one step further:
If particles were to be emitted from a static thermal source, 
and feed down corrections were performed, the ratio of resonance to
daughter particle would be independent of $m_\bot$.
That this ratio is in general  somewhat $m_\bot$ dependent is in this situation 
due to dynamical effects, such as hadronizing  matter flow and freeze-out geometry.
For this reason, it can be expected that the $m_\bot$ dependence of this ratio
can  help isolate these effects and thus  remove the ambiguities in
the present freeze-out models.

In Sec. II we  review  hadronization models and discuss their
ambiguities.  We  then show in Sec. III how resonance $m_\bot$ ratios
 can be used to distinguish between particle hadronization models. We close with a short discussion 
of open issues.

\section{Statistical hadronization}
\label{amb}
\subsection{General remarks }
Nearly all hadronic spectra comprise a significant {\cal O}(50\%) component from resonance decays.
Fits to data, which are not 
allowing for the decay contributions have a very limited  usefulness. 
Particle spectra and thus yields 
are in general, controlled by the properties of statistical hadronization 
models. However, some recent work fits the particle slopes
 only~\cite{burward-hoy,van-leeuwen,NA57,castillo}, 
treating the normalization of each particle
as a free parameter. 
This approach can be argued for assuming a long-lived posthadronization
``interacting hadron gas phase'' in which individual hadron abundances
subject to inelastic interactions evolve away from chemical equilibrium.
This particular  reaction picture clashes with e.g. the fact 
that  short-lived resonance ratios can be described 
within  the statistical hadronization 
 model using  the chemical (statistical hadronization) freeze-out 
 temperature obtained  in  stable particle studies \cite{PBMRHIC}.
This implies
 that in principle  the relative normalization  of the particle spectra 
should be derived from a hadronization scenario involving 
flavor chemical potentials. In fact a study of RHIC spectra 
 finds that the  normalization can be accounted for~\cite{florkowski}, and 
 that the chemical equilibration temperature also describes
particle spectra well. This is suggesting that  any posthadronization reinteraction
phase is short and has  minor influence on the particle yields.

The problem is that the different ways to derive hadronization
 particle distributions  have a profound effect on the resulting fitted temperature.
Temperature affects the absolute number of particles 
through several mechanisms  and anticorrelates 
with the  phase space occupancy parameters $\gamma_i, i=u,d,s$~\cite{ourspspaper,rafelski2002}.
It has been found that the introduction of these parameters, motivated by the need to conserve entropy at hadronization~\cite{rafelski2002} decrease
the $\chi^2/$ per degree of freedom considerably and lowers the freeze-out temperature by ~30 MeV
\cite{rafelski2002}.
Other workers assume
the light flavors are in chemical equilibrium~\cite{PBM99,becattini,PBM01,PBMRHIC}.

Considering ratios of  resonances to ground state  particles eliminates the fitted temperature's sensitivity to chemical equilibration, 
since the numerator and denominator have the same quark composition.
In every hadronization model considered here the chemical parameters cancel out
and only temperature and fireball freeze-out geometry and 
dynamics influence the observed ratios.

When fitting the particle spectra, the system's
spatial shape and the way the freeze-out progresses in time have
a considerable effect on the form of particle distributions, and hence
on the fitted temperature and matter flow.  
The impact of freeze-out geometry and dynamics on particle spectra were examined well before  RHIC data became available 
\cite{fireballspectra,fireballspectra2} and it was realized that
an understanding of freeze-out is essential for the statistical
analysis of the fireball \cite{heinzkolb}.  Even though this matter 
has been clearly recognized, a systematic
analysis  of how freeze-out geometry affects particle distributions
is for the first time attempted here. In fact, each of the models used in the study of particle
spectra \cite{van-leeuwen,ourspspaper,castillo,burward-hoy,florkowski}
employs  a different choice
of freeze-out geometry, based on different, often tacitly assumed, 
hadronization scenarios. Thus an understanding for the influence of hadronization
mechanism is impossible to deduce from this diversity. 

However, every study of firball hadronization  we are aware of  uses the  Cooper-Frye
formula~\cite{cooperfrye}:
\begin{equation}
\label{cooperfrye}
E \frac{d N}{d^3 p} = \int p^{\mu} d^3 \Sigma_{\mu} f (p^{\mu} u_{\mu}, T, \lambda) 
\theta(p^{\mu} d^3 \Sigma_{\mu}),
\end{equation}
where $p^{\mu}$ is the particle's four-momentum, $u^{\mu}$ is the systems velocity profile, \textit{T} is the temperature,
$\lambda$ is a chemical potential, $f(E,T,\lambda)$ is the statistical distribution of the emitted particles in terms of
energy and conserved quantum numbers
and $\Sigma^{\mu}$ describes the hadronization geometry.  It is the covariant
generalization of a volume element of the fireball, i.e., a
 ``3D'' surface in spacetime from which particles are emitted.
$\theta(p^{\mu} d^3 \Sigma_{\mu})$ is the step function, which
eliminates  the possible inward emission~\cite{cf_trunc1,cf_trunc2}.

The Cooper-Frye formula, Eq. (\ref{cooperfrye}), is believed to be 
the most general way to implement statistical hadronization
emission of particles.
For it to represent a physical description of the system, the following two conditions
have to be met:\\
\indent (i) Statistical hadronization must apply.
The particles emitted from a volume element (in it's co-moving
frame), will be distributed according to
the Bose-Einstein or Fermi-Dirac distributions $f (E, T, \lambda)$
for some temperature $T$ and fugacity $\lambda$.\\
\indent (ii) 
A ``small'' volume element hadronizes rapidly in it's rest frame, that is, 
no long lived mixed Quark gluon plasma (QGP) -hadronic confined phase exists.

If this second condition is satisfied, it becomes possible to define
a hadronization hypersurface $\Sigma^{\mu}=\mathbf{(}t_f(x,y,z),x,y,z \mathbf{)})$ which specifies
at which  time $t_f$ hadrons are emitted from the point $(x,y,z)$.
In this fast hadronization case differing $\Sigma^{\mu}$
can be considered for  physically differing  models.   
The different choices of $\Sigma^{\mu}$  correspond to physically
different scenarios, and it becomes possible, in principle, to
distinguish them experimentally.
However, if a long-lived  mixed phase does exist,
 it might well be that the Cooper-Frye
formula can be used as an approximation technique to transform
a hydrodynamically evolving system into hadrons and authors who worked with 
a long mixed phase have chosen this approach, see e.g. Ref. ~\cite{shuryak}.

\subsection{Freeze-out geometry}
The high baryon stopping power observed at SPS energies
\cite{allNA49,piNA49,lamNA49} has prompted some authors to use a
spherical expansion and freeze-out as an ansatz
\cite{ourspspaper}.

However, at RHIC collision energies the measured 
$dN/d \eta$ ~\cite{phobos,brahms}
indicates that around mid-rapidity the system conditions can be approximated
by the Bjorken picture~\cite{bjorken}.

To describe particle spectra measured around midrapidity, therefore, 
boost invariance 
becomes the dominant symmetry on which freeze-out geometry should be based.
To construct such a hadronization scenario, we consider that the most general
cylindrically symmetric flow profile
\begin{equation} 
u^{\mu} = 
\left( 
\begin{array}{l}
\cosh (y_L) \cosh(y_\bot) \\ 
\sinh(y_\bot) \cos(\theta) \\ 
\sinh(y_\bot) \sin(\theta) \\ 
\sinh (y_L) \cosh(y_\bot) 
\end{array} 
\right),\quad
p^{\mu} = 
\left( 
\begin{array}{c}
m_\bot \cosh (y)  \\ 
p_\bot \cos(\phi) \\ 
p_\bot \sin(\phi) \\ 
m_\bot \sinh (y)  
\end{array} 
\right)
\label{flowprof}
\end{equation}
 (the last, longitudinal
 coordinate is defined along the beam direction) 
leads to the following rest energy~\cite{rafrap}:
\begin{equation}
\label{goesas}
p_{\mu} u^{\mu} = m_\bot \cosh(y_\bot) \cosh(y-y_L) - \sinh(y_\bot)  \cos(\theta-\phi) p_\bot.
\end{equation}
The requirement for the Bjorken picture is that the emission volume element
has the same $y_L$ dependence:
\begin{equation}
p_{\mu} d^3 \Sigma^{\mu} \sim A \cosh(y-y_L)+B.
\end{equation}
This constrains the freeze-out hypersurface to  be
of the form
\begin{equation}
\label{bjorkfreeze}
\Sigma^{\mu}= (t_f \cosh(y_L),x,y,t_f \sinh(y_L)).
\end{equation}
Here $t_f$ is a parameter invariant under boosts in the \textit{z} direction,
whose physical significance depends on the model considered.

For central collisions, a further simplifying constraint
is provided by the cylindrical symmetry, which forces $t_f$,
 as well as $y_L$ and $y_\bot$, to be independent of the angles
$\theta$ and $\phi$.
The freeze-out hypersurface can be parametrized, in this case, as
\begin{equation}
\label{cylfreeze}
\Sigma^{\mu}=\left( t_f(r) \cosh(y_L),r \sin(\theta),r \cos(\theta),t_f (r) \sinh(y_L) \right),
\end{equation}

\begin{eqnarray}
\label{blastsurf}
d^3 \Sigma^{\mu} = t_f r dr d \theta d y_L \nonumber \\
\left(
\cosh(y_L) 
 \frac{\partial t_f}{\partial r} \cos(\theta),
 \frac{\partial t_f}{\partial r} \sin(\theta),
\sinh(y_L)
\right)
\end{eqnarray}
And the emission element takes the form
\begin{eqnarray}
\label{blastemission}
p^{\mu} d^3 \Sigma_{\mu} =\nonumber \\ \left[ m_\bot \cosh(y-y_L)
 - p_\bot \frac{\partial t_f}{\partial r} \cos(\theta-\phi) \right] t_f r dr d \theta d y_L,
\end{eqnarray}
with the same dependence on the angle as Eq. (\ref{goesas}).
Equation (\ref{cooperfrye}) can then 
be integrated over all the possible values
of $y_L$ and $\theta-\phi$ to give a particle spectrum depending purely
on the transverse mass, temperature, and $y_\bot$.
The fits in~\cite{burward-hoy,van-leeuwen,castillo,florkowski}
are based on such an ansatz.

What distinguishes the models currently considered is the time component
of the freeze-out surface.
The most general freeze-out hypersurface compatible with cylindrical 
symmetry is
provided by Eq.(~\ref{cylfreeze}). Generally, $t_f$ (a generic
function of \textit{r}) represents the time, in a frame comoving with the
longitudinal flow, at which the
surface at distance \textit{r} freezes out.

The fits in Refs. ~\cite{burward-hoy,van-leeuwen,castillo} are based on a particular
case of such a freeze-out surface, in which $t_f$ is completely independent
of $r$ ($\partial t_f/\partial r=0$).
Such a picture's  physical reasonableness can be questioned, e.g., 
why should spatially distant volume
elements, presumably with different densities and moving at different 
transverse velocities, all freeze out simultaneously in a longitudinally
comoving frame?. However, such a simple model  can perhaps serve as an 
 approximation.

More generally, the ``burning log'' model  \cite{fireballspectra,blastwave}
(sometimes referred to as ``blast wave'';
This term, however, is also used to refer to the 
$\partial t_f/\partial r=0$ model described in the preceding paragraph) assumes that the 
emission occurs through a
three-dimensional hadronization surface which is moving 
at a constant ``velocity'' ($1/v_f=\partial t_f/\partial r$ throughout
the fireball. Both boost-invariant and spherically symmetric versions 
of burning log model were considered. Even if the hadronization
velocity encompasses an extra parameter  $v_f$, it is worth
considering since it is based on a  physically motivated hadronization  picture.
Moreover, the burning log picture is a suitable framework in the study 
of sudden hadronization. Sudden hadronization occurs
when the fireball encounters a mechanical instability~\cite{sudden},
which combined with the fireball's high transverse
flow ensures that the emission surface spreads to the interior of the
fireball with $v_f\simeq c$.
All of the indications suggested for such a picture seem to be borne out
by both SPS and RHIC data~\cite{rafelski2002,sudden,pratt}.

\begin{figure*}
\centerline{\resizebox*{!}{0.3\textheight}
{
\includegraphics{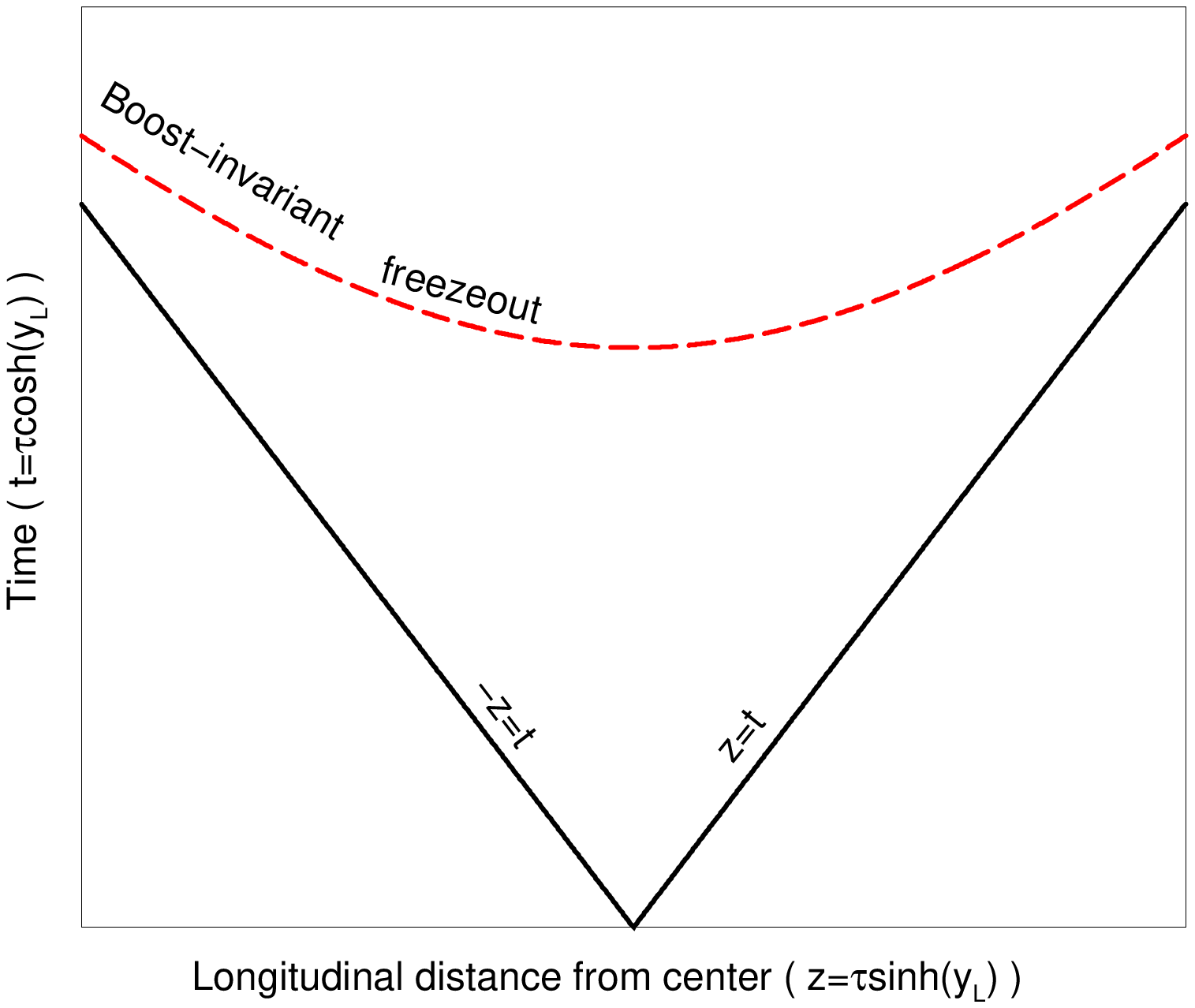}
}
\resizebox*{!}{0.3\textheight}
{
\includegraphics{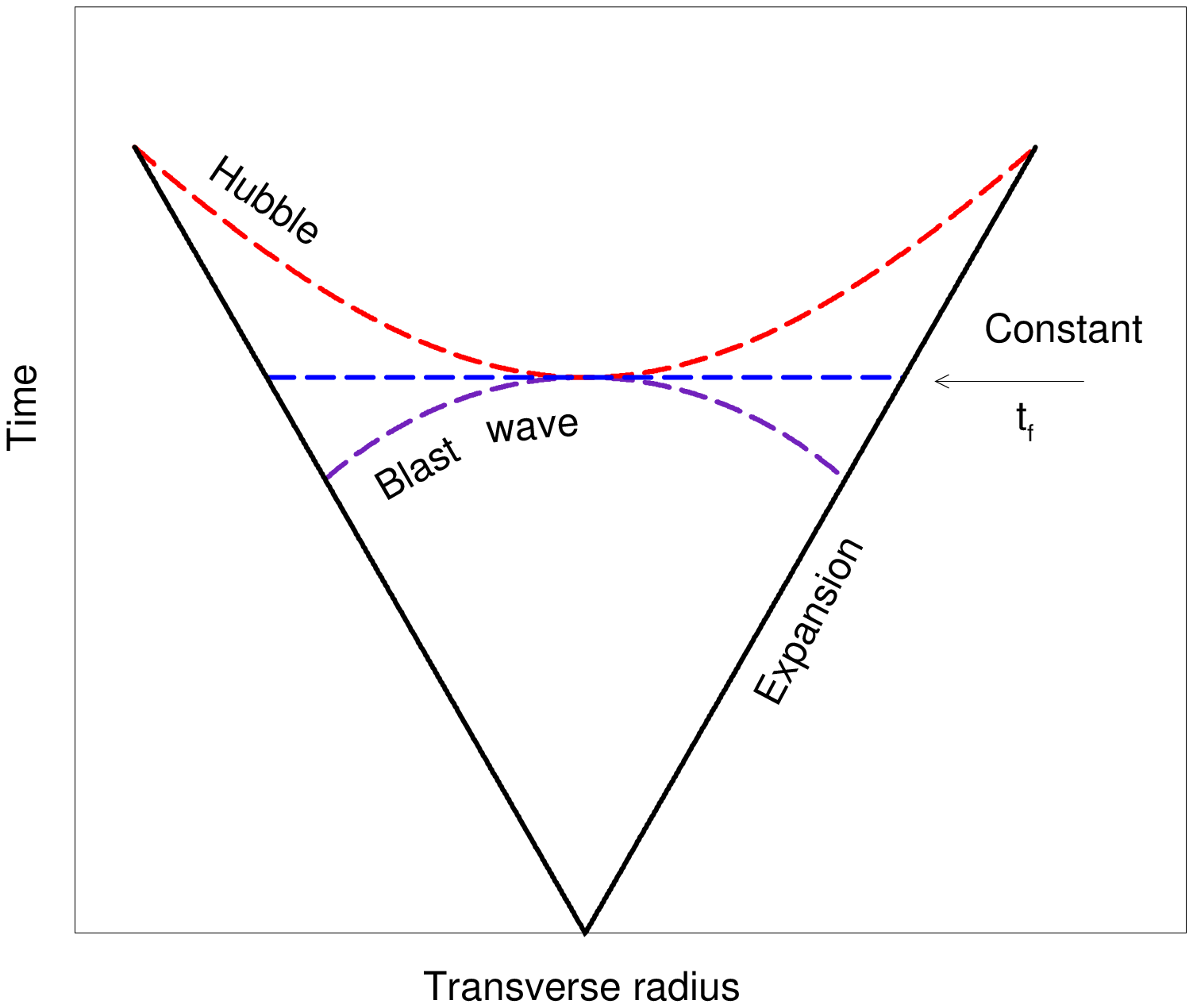}
}}
\caption{(Color online) While boost-invariance fixes the longitudinal
freeze-out structure (left), several scenarios exist for the transverse dependence of
 freeze-out (right).   For spherical freeze-out, only plot on the right applies 
\label{figsurf}}
\end{figure*}

\begin{table*}
\caption{(Color online) Freeze-out hypersurfaces at contours of
  constant radii. \label{hypsurf}}
\vspace{0.2cm}
\begin{ruledtabular}
\begin{tabular}{lccr}
\hline
Surface 
& 
$\Sigma^{\mu}$ 
&  
$E \frac{dN}{dp^3}$ \footnote{$\beta=\cosh(y_\bot)/ T ,\alpha=\sinh(y_\bot)/T$}
 & reference
\\ \hline
$\begin{array}{c}
\mbox{Constant\ } t_f \\
{\partial t_f}/{\partial r}=0
\end{array}
$
& 
$\left( \begin{array}{c}
t_f \\ \vec{r} \end{array} \right)$ 
& 
$m_\bot K_1 (\beta m_\bot)I_{0} (\alpha p_\bot)$
&
~\cite{NA57,van-leeuwen,burward-hoy,castillo,keranen}
\\ \hline
$\begin{array}{c}
\mbox{Hubble}\\
(\mbox{constant}\ \tau_f) 
\end{array}
$
& 
$\tau_f 
\left( \begin{array}{l}\cosh(y_L) \cosh(y_\bot) \\ 
       \sinh(y_\bot) \cos(\theta) \\
        \sinh(y_\bot) \sin(\theta) \\ 
          \sinh(y_L) \cosh(y_\bot) 
\end{array} 
\right)
$ 
&
$\begin{array}{c}
m_\bot \cosh(y_\bot) I_{0} (\alpha p_\bot) K_1 (\beta m_\bot )-\\ 
p_\bot \sinh(y_\bot) I_{1}(\alpha p_\bot)  K_0 (\beta m_\bot)
\end{array}
$
& 
\cite{florkowski}
\\ \hline
$\begin{array}{c}
\mbox{Blast/burning log}\\
\mbox{(boost invariant)}
\end{array}
$
& 
$\left( \begin{array}{c} t_f (r) \cosh(y_L) \\
                  r \cos(\theta)\\        
                  r \sin(\theta)\\ 
                  t_f (r) \sinh(y_L) \end{array} \right)$
& 
$\begin{array}{c}
m_\bot  I_{0} (\alpha p_\bot)  K_1 (\beta m_\bot)
-\\
p_\bot \frac{\partial t_f}{\partial r} I_{1} (\alpha p_\bot)    K_0 (\beta m_\bot)
\end{array}
$
&
This paper,~\cite{florkowski}
\\ 
\hline 
\\
$\begin{array}{c}
\mbox{Blast/burning log}\\
(\mbox{spherical})
\end{array}
$
& 
$\left( \begin{array}{c} t_f \\ r \vec{e}_r \end{array} \right)$
&
$\begin{array}{c}
e^{-E/T} \sqrt{\frac{T}{p_\bot \sinh(y_\bot)}} 
(E I_{1/2} (\alpha p_\bot) -\\
 p_\bot \frac{\partial t}{\partial r} I_{3/2} (\alpha p_\bot))
\end{array}$
&
\cite{ourspspaper,fireballspectra}
\\ \hline
\end{tabular}
\end{ruledtabular}
\end{table*}

An approach based on the hypothesis of initial state ``synchronization''
by the primary instant of collision and the following independent but 
equivalent evolution of all volume elements  assumes that each element 
of the system undergoes freeze-out at the same proper time $\tau$.
In this framework 
each fireball element expands and cools down
independently, hadronizing
when it's temperature and density reach the critical value.
This model  was successfully used to describe 
RHIC $m_\bot$-spectra~\cite{florkowski}.
In this approach  $t_f$ in Eq. (\ref{cylfreeze}) is equal to
$\tau \cosh(y_\bot)$ and the hadronization hypersurface in Eq. (\ref{blastsurf})
becomes proportional to the flow vector:
\begin{eqnarray}
\label{hubblefreeze}
\Sigma^{\mu}&=& \tau u^{\mu}\\
d^3 \Sigma^{\mu} &=& \tau r dr d\theta d y_L u^{\mu}=dV u^{\mu}\\
r &=& \tau \sinh(y_\bot).
\end{eqnarray}
In this   hadronization model  the heavy ion fireball  behaves similarly
to the expanding Hubble universe. In the  
`Hubble' scenario, the Cooper-Frye formula
reduces to the Touscheck Covariant Boltzmann distribution
~\cite{rafrap,jansbook,tous1,tous2}.
\begin{eqnarray}
\label{touscheck}
\frac{V_0d^3p}{(2\pi)^3} e^{-E/T}&\to& \frac{V_\mu p^\mu}{(2\pi)^3}
d^4p\,2\delta_0(p^2-m^2)e^{-p_\mu u^\mu/T}\, \\
V^{\mu}&=&V_0 u^{\mu}
\end{eqnarray}
(Where \textit{V} is the co moving fireball's volume element in the local
rest frame.)

To summarize and illustrate the diversity of distinct  hadronization 
geometries we present in Table~\ref{hypsurf} and Fig.~\ref{figsurf} the 
freeze-out scenarios examined here. As we shall see the 
 choice of freeze-out geometry produces in a fit of experimental 
data a non trivial
effect capable of altering significantly the understanding 
of statistical hadronization parameters.

\begin{figure*}
\centerline{\resizebox*{!}{0.3\textheight}
{
\includegraphics{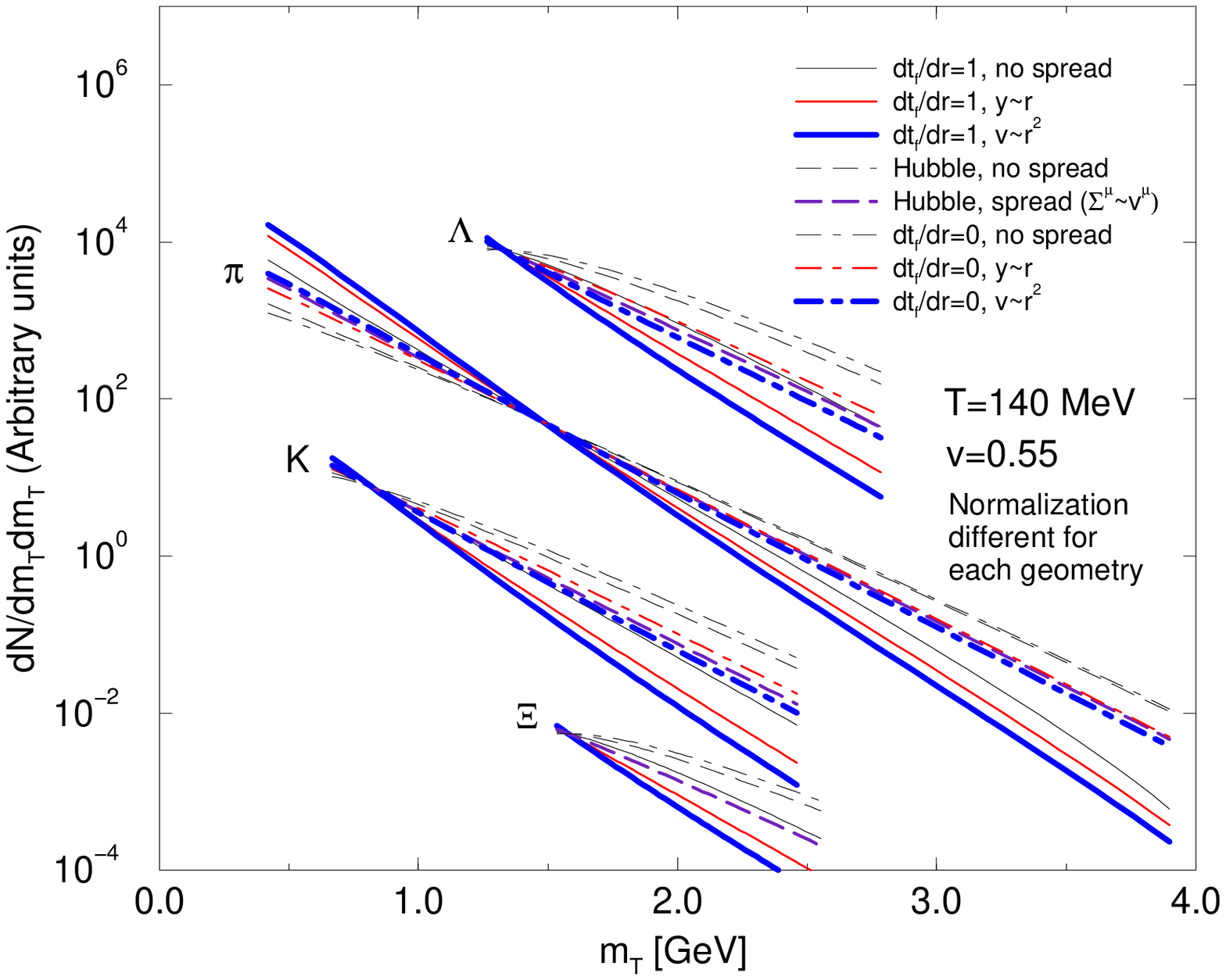}
}
\resizebox*{!}{0.3\textheight}{
\includegraphics{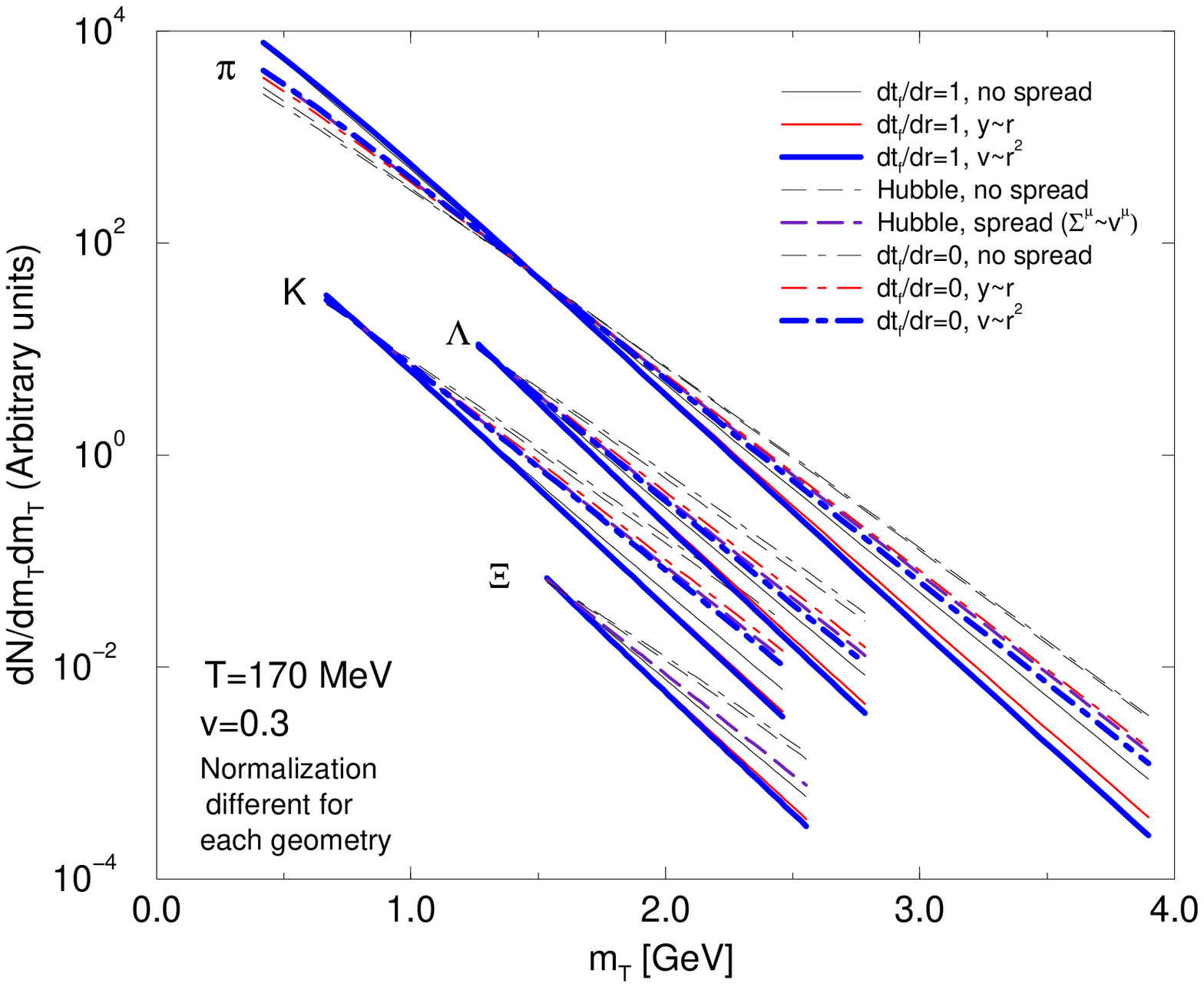}
}}
\caption{(Color online) $\pi$,$K$,$\Lambda$ and $\Xi$ $m_\bot$ distributions 
obtained with different freezeout models and flow profiles.
For this and subsequent figures, a uniform density profile was assumed
\label{mcplots}}
\end{figure*}

\subsection{Flow profile}
Hydrodynamical expansion of the fireball implies in 
general that  each volume element
 will have a different density and transverse expansion rate.
For this reason, the integral over $d^3 \Sigma$ can  span a range
of flows, weighted by density. In first approximation one can  
fit data using just an ``average'' flow
velocity throughout the entire fireball~\cite{NA57,ourspspaper}:
\begin{eqnarray}
E \frac{dN}{d^3 p}&=& \int r dr (E-p_\bot \frac{dt_f}{dr}) f(T,y_\bot (r),\lambda) 
\nonumber\\
&\propto& (E-p_\bot \frac{dt_f}{dr}) f(T,\langle y_\bot\rangle,\lambda).
\end{eqnarray}

However, if one wants to properly identify 
$\frac{dt_f}{dr}$, the  flow profile should be  taken into account~\cite{comingpaper}.
Hydrodynamic simulations~\cite{shuryak} accompanied by assumption that
 freeze-out happens when a volume element reaches a critical
energy density indicate that the transverse rapidity will depend linearly
with the radius i.e. $v_\bot \sim \tanh(r)$.
This condition, however, is appropriate for a static freeze-out
and will not in general hold if the freeze-out is sudden.
Other flow profiles have been tried in the literature, arising 
from dynamical hypothesis. For example, 
the assumption that the  freeze-out occurs at the same time
$t_f$ results in a quadratic ($v \propto r^2$) flow profile~\cite{hydroheinz}, 
which has also  been used recently in fits to data~\cite{castillo}.
In the  Hubble fireball~\cite{florkowski} the  freeze-out conditions
will also result in a distinctive flow profile. Specifically with 
$\Sigma^{\mu} \propto u^{\mu}$, we have $\gamma v \propto r$.

Density profiles  also depend on the assumed initial condition
and the equation of state of the expanding QGP.
It has been shown~\cite{wa98} that different density choices have
a considerable effect on both the temperature and flow fits at SPS
energies.  

Figure.~\ref{mcplots} shows how the choice in
hadronization dynamics and flow profiles at same  given
freeze-out temperature and transverse flow can result in a range 
of inverse spectral slopes. Here the density profiles were assumed to be uniform.
It is  clear that the same freeze-out parameters  give
rise to a variety of substantially different particle
spectra. Conversely, fits to experimental data will only produce 
reliable information on the freeze-out conditions if and when we
have a prior knowledge of the hadronization geometry and dynamics. 
Therefore, conclusions about statistical model fits,
as well as arguments whether freeze-out occurs  simultaneously for different particles
or not, cannot be answered  while the models used to fit the data are plagued by
such uncertainties. We will  now turn to study how the measurement  of 
spectra of short-lived resonances might provide us with a way of making
progress.

\section{Momentum dependence of the resonance-particle ratios as a freeze-out probe}
\label{res}
We have shown that the measurement of resonances
can probe both the hadronization temperature, and
the lifetime of the interacting hadron gas phase
\cite{sqm2001,resonances}.
Ratios of a generic resonance (henceforward called $Y^*$) to the 
light particle (which we will refer to as $Y$) with an identical
number of valence quarks are particularly sensitive  probes of freezeout temperature
because chemical dependence cancels out within the ratio. 
If we examine this ratio within a given  $m_\bot>m_{Y^*}$ range,
 we expect to disentangle flow and freeze-out conditions,
since the ratio $Y^*/Y$ should not depend on $m_\bot$ for a purely static and 
 thermal source.

We therefore take the most general Boost-invariant freeze-out hypersurface
in the Boltzmann limit (see Table~\ref{hypsurf}.  Boost invariant
implies this is a good approximation at midrapidity)
\begin{equation}
\label{blastgen}
\frac{dN}{dm_{T}^2} \propto S(m_\bot,p_\bot)=\int_{\Sigma} r dr \mathcal{S} (m_\bot,p_\bot,r),
\end{equation}
where
\begin{eqnarray}
\label{sform}
\mathcal{S}(m_\bot,p_\bot,r) &=&\nonumber 
m_\bot K_1 (\beta m_\bot) I_{0} (\alpha p_\bot)\\
&-& \frac{\partial t_f}{\partial r} p_\bot K_0 (\beta m_\bot) I_1
 (\alpha p_\bot),
\end{eqnarray}
with
\begin{eqnarray}
\label{alphabeta}
\beta  = \frac{\cosh[y_\bot(r)]}{T},\quad 
\alpha = \frac{\sinh[y_\bot(r)]}{T}
\end{eqnarray}
and use it to calculate the ratio between two particles with the same
chemical composition.
The chemical factors cancel out, and we are left with
\begin{equation}
\label{blastratio}
\frac{Y^*}{Y} = \left(\frac{g^*}{g}\right) 
\frac{S(m_\bot,p_\bot^*)}{S(m_\bot,p_\bot)},
\end{equation}
where $g^*$ and $g$ refer to each particle's degeneracy and
the function $S(m_\bot,p_\bot)$ is given by Eq.~(\ref{sform}).
(Note that $m_\bot$ is the same for $Y^*$ and $Y$, but $p_\bot$ varies).

\begin{figure*}
\centerline{\resizebox*{!}{0.32\textheight}{
\includegraphics{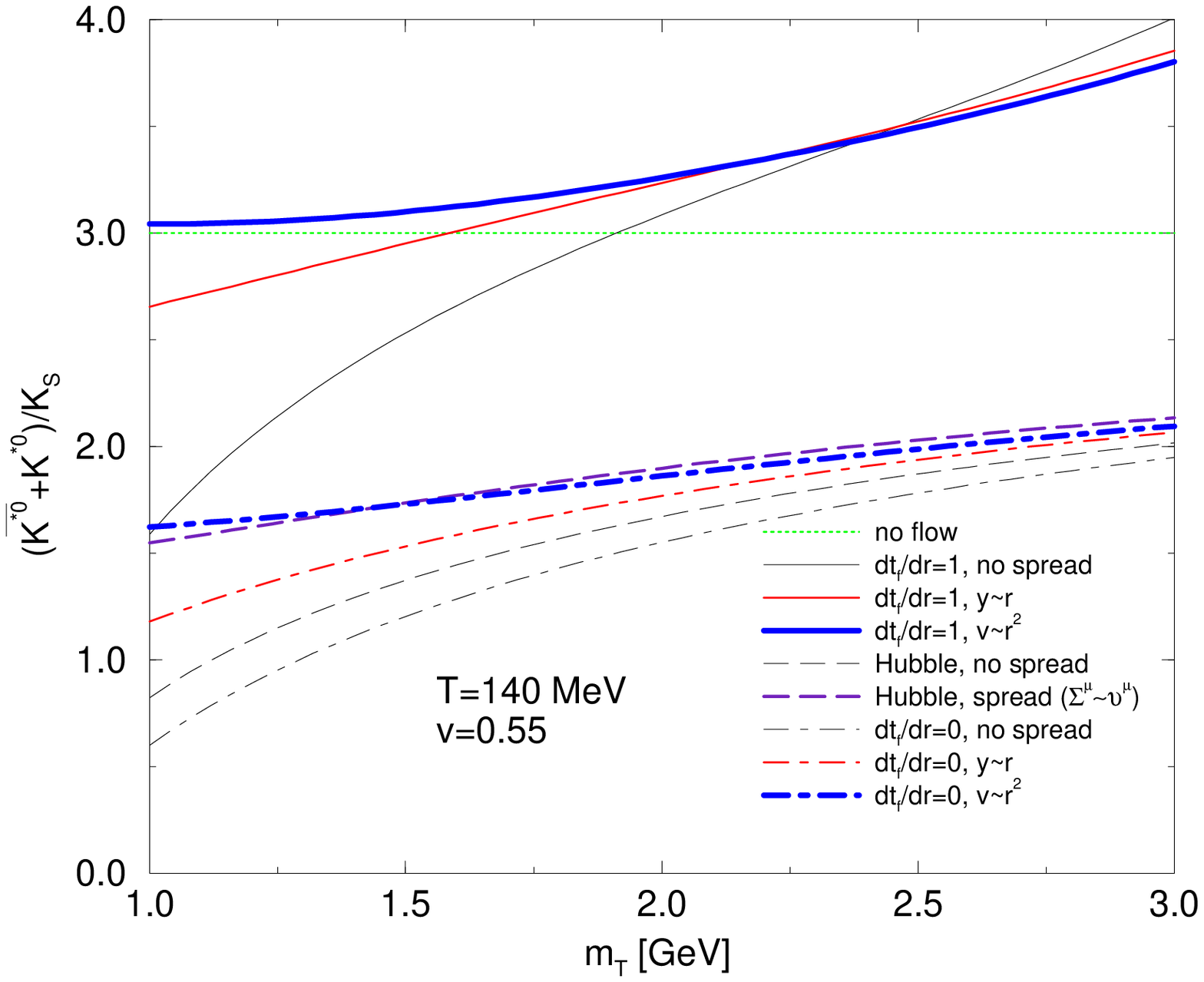}
}
\resizebox*{!}{0.32\textheight}{
\includegraphics{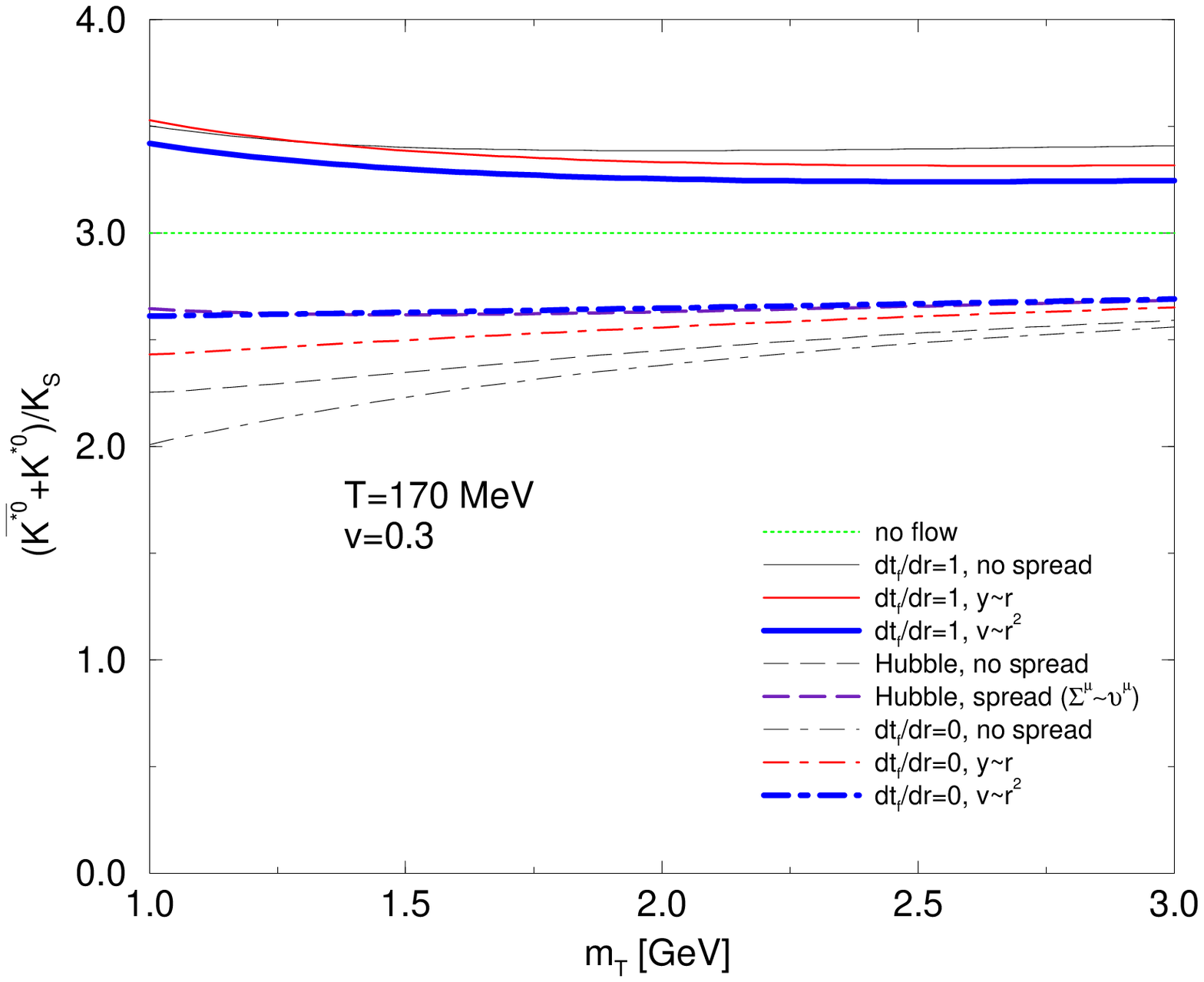}
}}
\centerline{\resizebox*{!}{0.32\textheight}{
\includegraphics{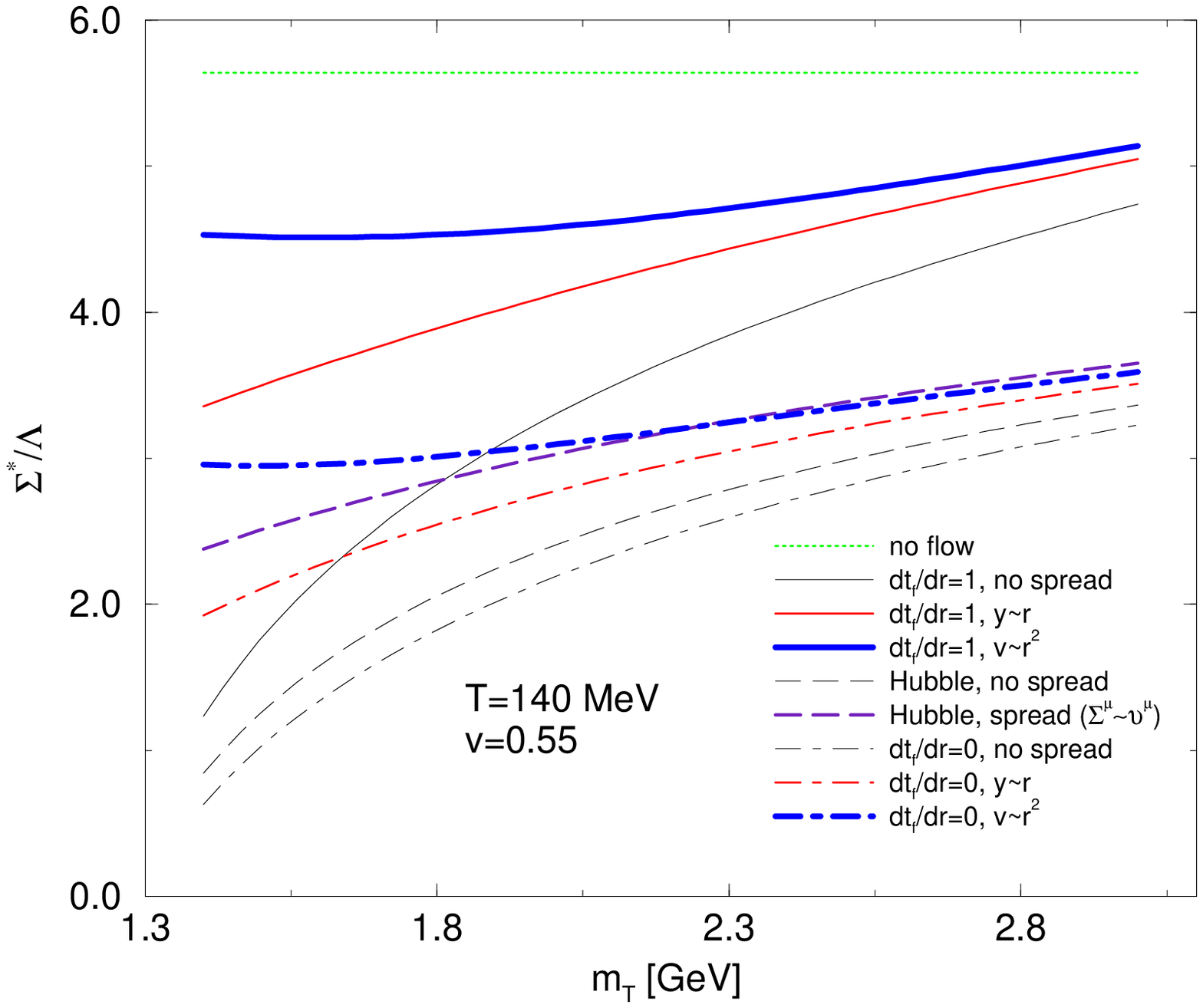}
}
\resizebox*{!}{0.32\textheight}{
\includegraphics{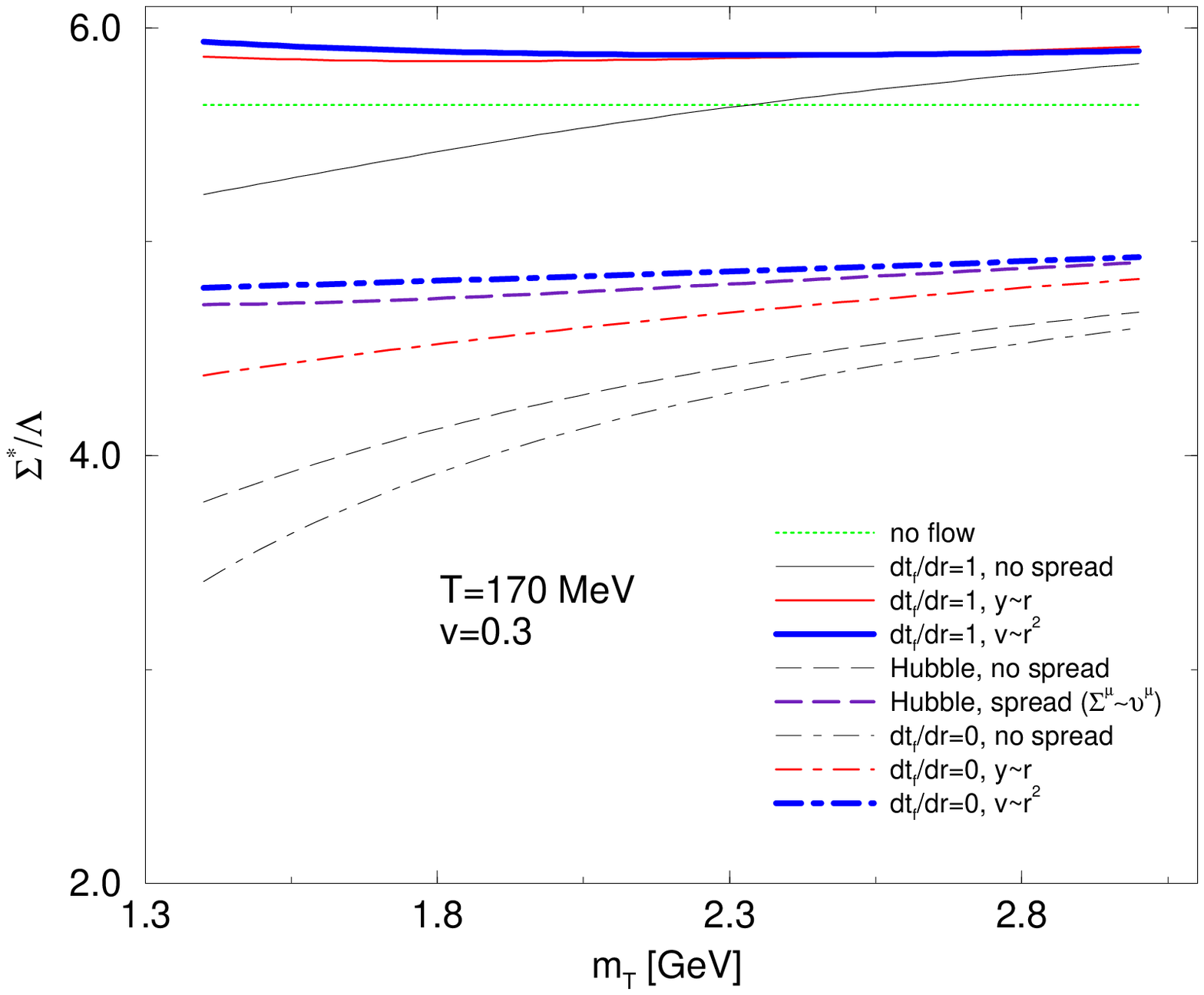}
}}
\centerline{\resizebox*{!}{0.32\textheight}{
\includegraphics{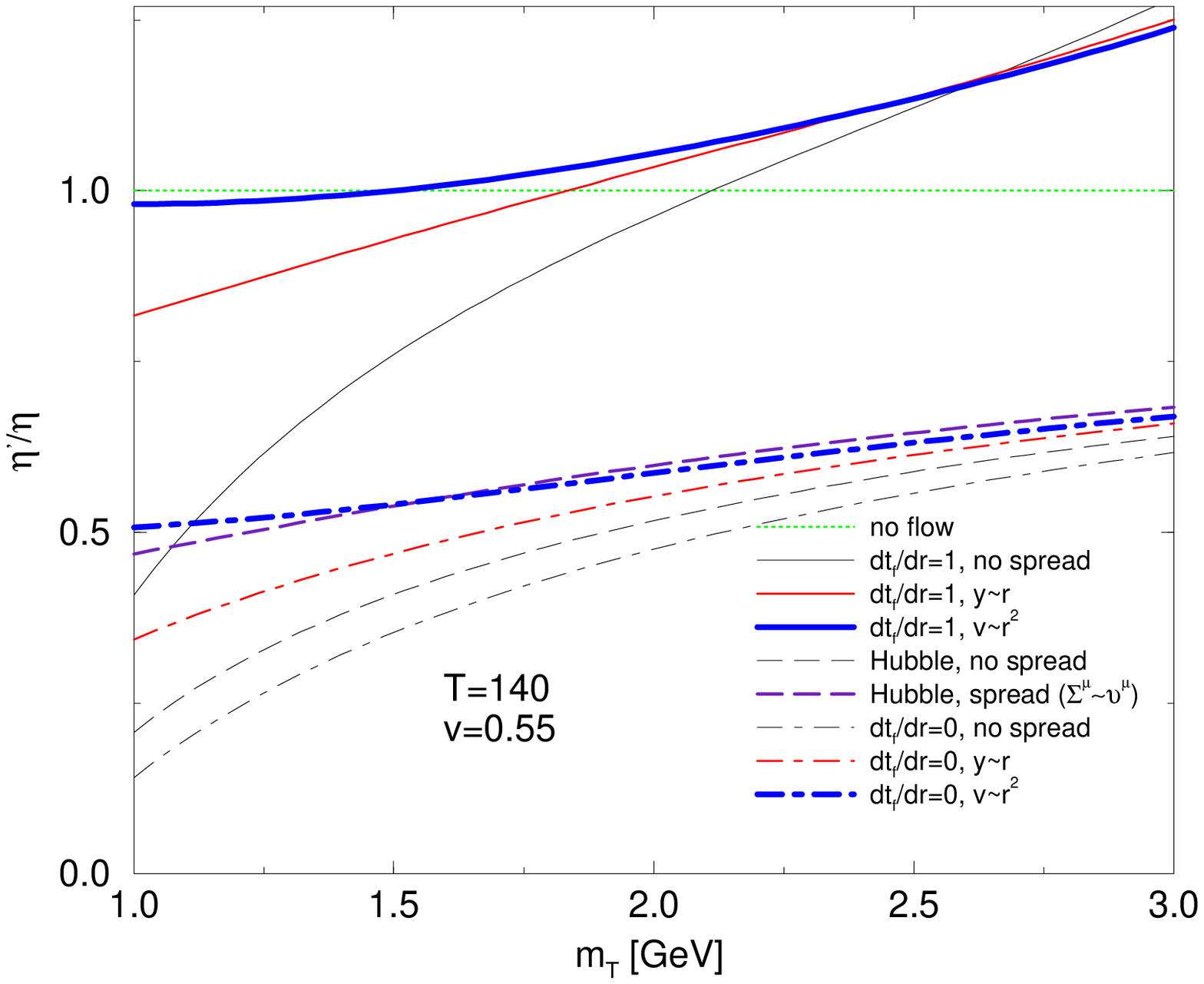}
}
\resizebox*{!}{0.32\textheight}{
\includegraphics{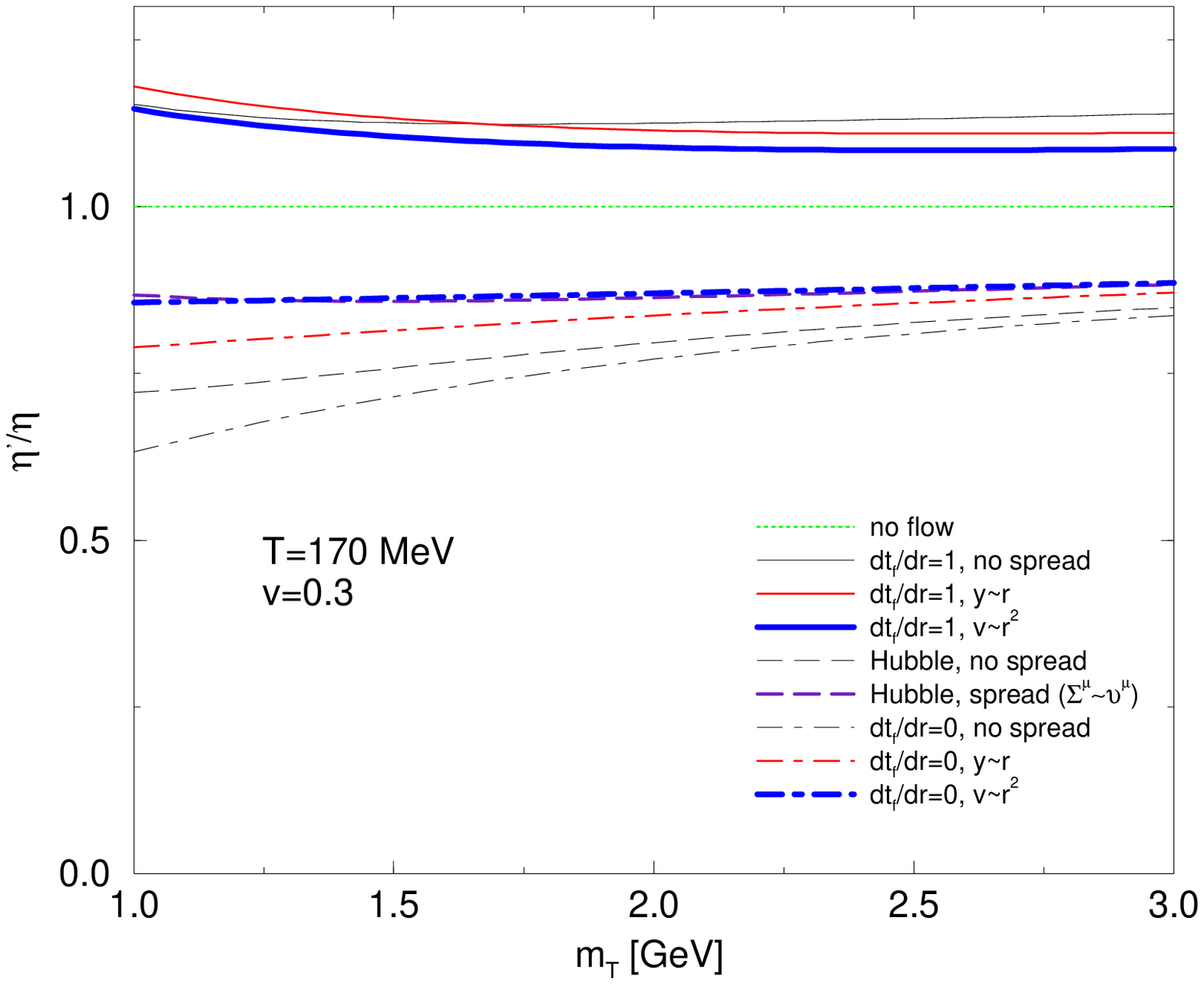}
}}
\caption{(Color online) Dependence of the  $K^*/K$, $\Sigma^*/\Lambda$ and $\eta'/eta$ on
the Freeze-out model  \label{diagres}.}
\end{figure*}

Figure ~\ref{diagres}  shows the application of this procedure
to the cases  $(K^*+\overline{K^*})/(K_S)$ (top), $\Sigma^*(1385)/\Lambda$ (middle), and
$\eta'/\eta$ (bottom) at two freeze-out temperatures and flows:
$T=140 \mbox{MeV}, v_{max}/c=0.55$ on left and
$T=170 \mbox{MeV}, v_{max}/c=0.3$ on the right.  Significant deviations 
from  simple constant values are observed, showing the sensitivity of the
ratio to freeze-out geometry and dynamics. 
The analytically simple result  in Eq.~(\ref{blastratio}) is
 valid only if the light particle $Y$ has
been corrected for feed down from resonances, including $Y^*$.
In other words, Eq.~(\ref{blastratio}) as well
as Fig.~\ref{diagres} require that decay products from reconstructed
$Y^*$ do not appear on the bottom of the ratio.
Experiments usually do not
do such feed down corrections~\cite{fachini,vanburen,markert,friese},
 since this would
increase both statistical and systematic error on the ratio, and 
it is not always possible
to do such corrections at all (undetected decays) 
 or in the full range of experimental sensitivity.

Introducing the feed down corrections into Eq.~(\ref{blastratio}),
we obtain
\begin{equation}
\label{blastratiofeed}
\frac{Y^*_{observed}}{Y_{observed}} =\frac{g^* S(m_\bot,p_\bot^*)}{g S(m_\bot,p_\bot) + \sum_{i} g^{*}_{i} b_{Y^*_i \rightarrow Y} R(m_\bot,p_{Ti})  }.
\end{equation}
Here, $S(m_\bot,p_\bot)$ describes the directly
produced particles and has the form given by Eq.~(\ref{blastgen})
and each term $R(m_\bot,p_{Ti}^{*})$ describes a feed down contribution.

In the case of an incoherent many-particle system, such as that
we are dealing with, the dynamical (matrix element) part of the decay
amplitude factors out~\cite{respectra}, and
$R(m_\bot,p_{Ti}^{*})$ is obtained by integrating the 
statistical hadronization distribution with a weight given by the phase
space elements of the decay products.
Thus, for a generic $Y^* \rightarrow Y$ feed down is given by an $N$-body decay,
\begin{eqnarray}
\label{rform}
R(m_\bot,p_\bot) &=& \\ \nonumber
&&\hspace*{-1cm}\int \prod_{j=2}^{N} \frac{d^3
p_{j}}{E_{j}}
S(m_{T}^{*},p_{T}^{*})
\delta \left( p_{\mu}^{*}-p_{\mu}-\sum_{2}^{N} p_{j \mu} \right),
\end{eqnarray}
where the integral is performed over the whole   allowed region.
If more than one feed down occurs, Eq.~(\ref{rform}) can be used iteratively, 
with the left-hand side to be fed back to the right-hand side at each successive iteration.

In general, this expression can get very complicated, and the Monte Carlo
integration becomes necessary.   For most cases considered here, where
there is one feed down and two or three body decays, Eq. \ref{rform} can be
carried out semianalytically
\cite{florkowski,resonances,fireballspectra}.

\begin{figure*}
\begin{center}
\centerline{\resizebox*{!}{0.32\textheight}{
\includegraphics{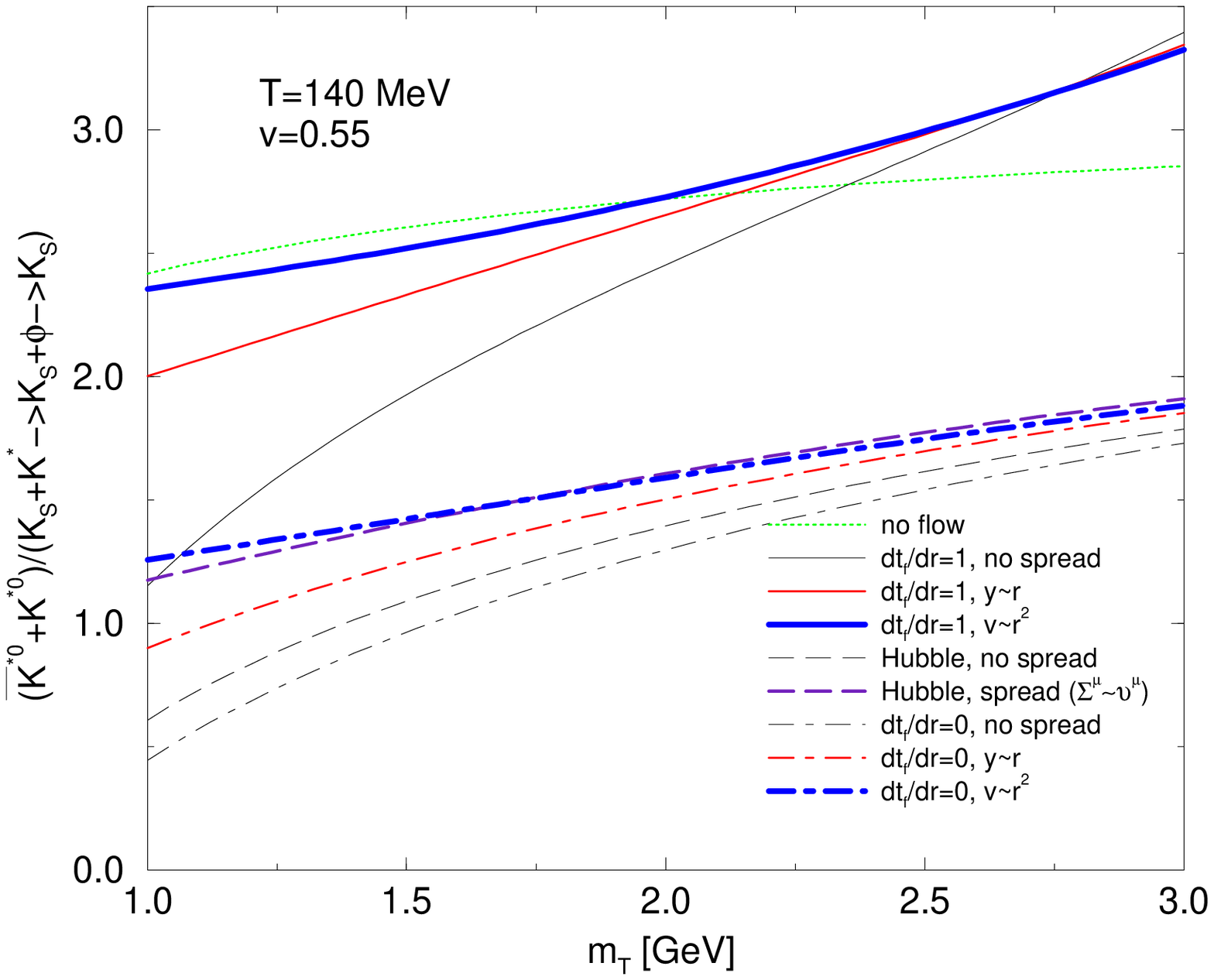}
}
\resizebox*{!}{0.32\textheight}{
\includegraphics{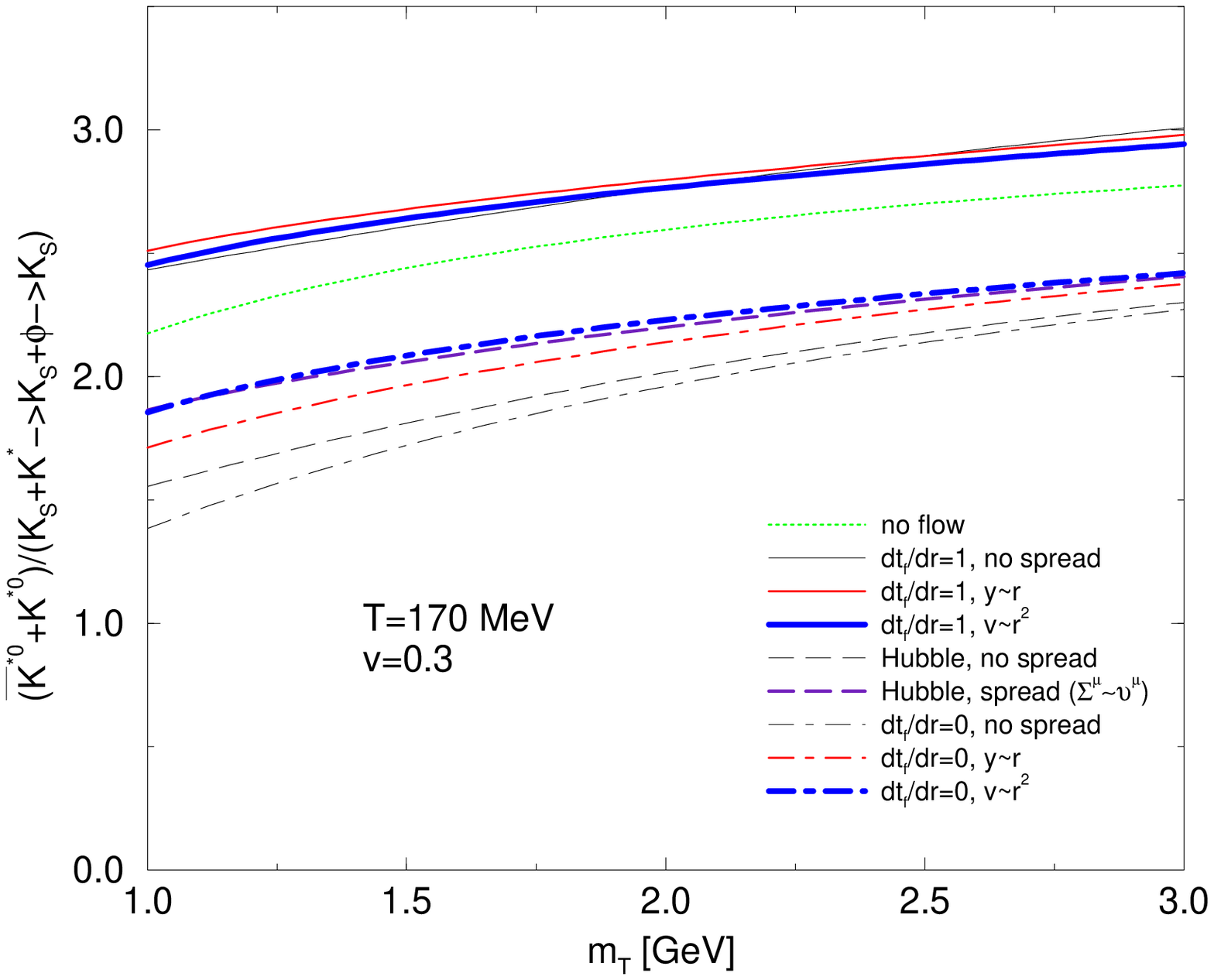}
}}
\centerline{\resizebox*{!}{0.32\textheight}{
\includegraphics{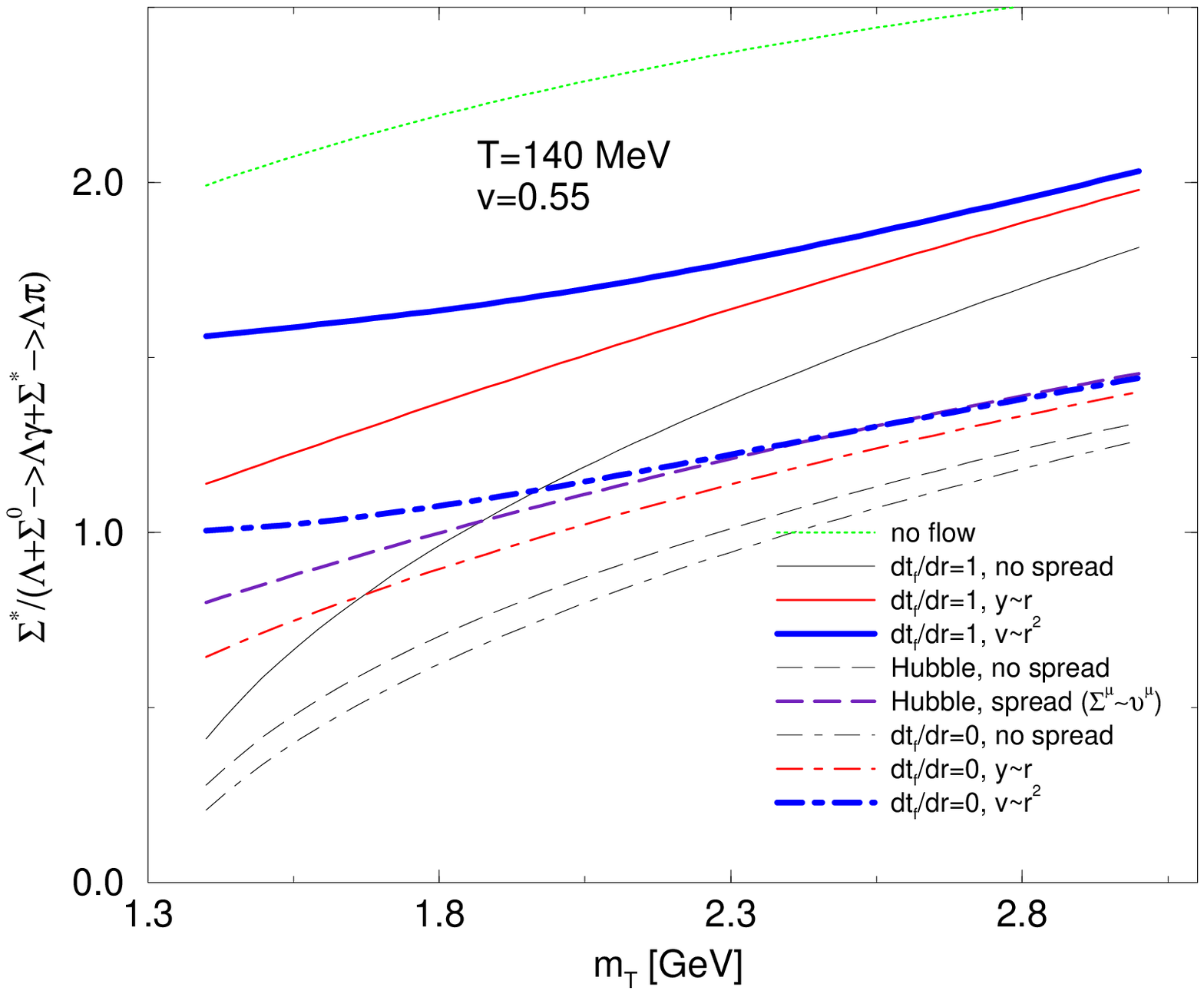}
}
\resizebox*{!}{0.32\textheight}{
\includegraphics{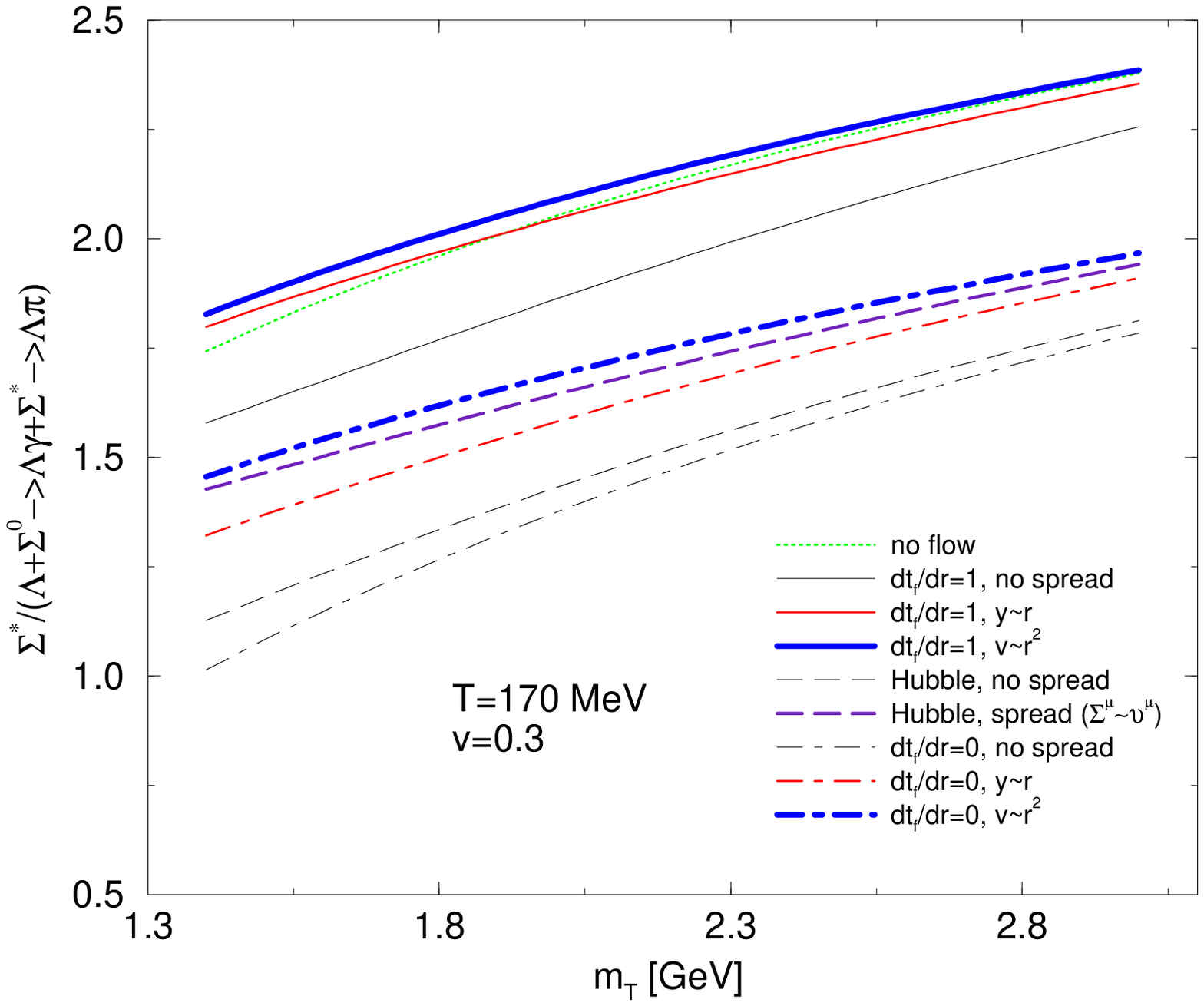}
}}
\centerline{\resizebox*{!}{0.32\textheight}{
\includegraphics{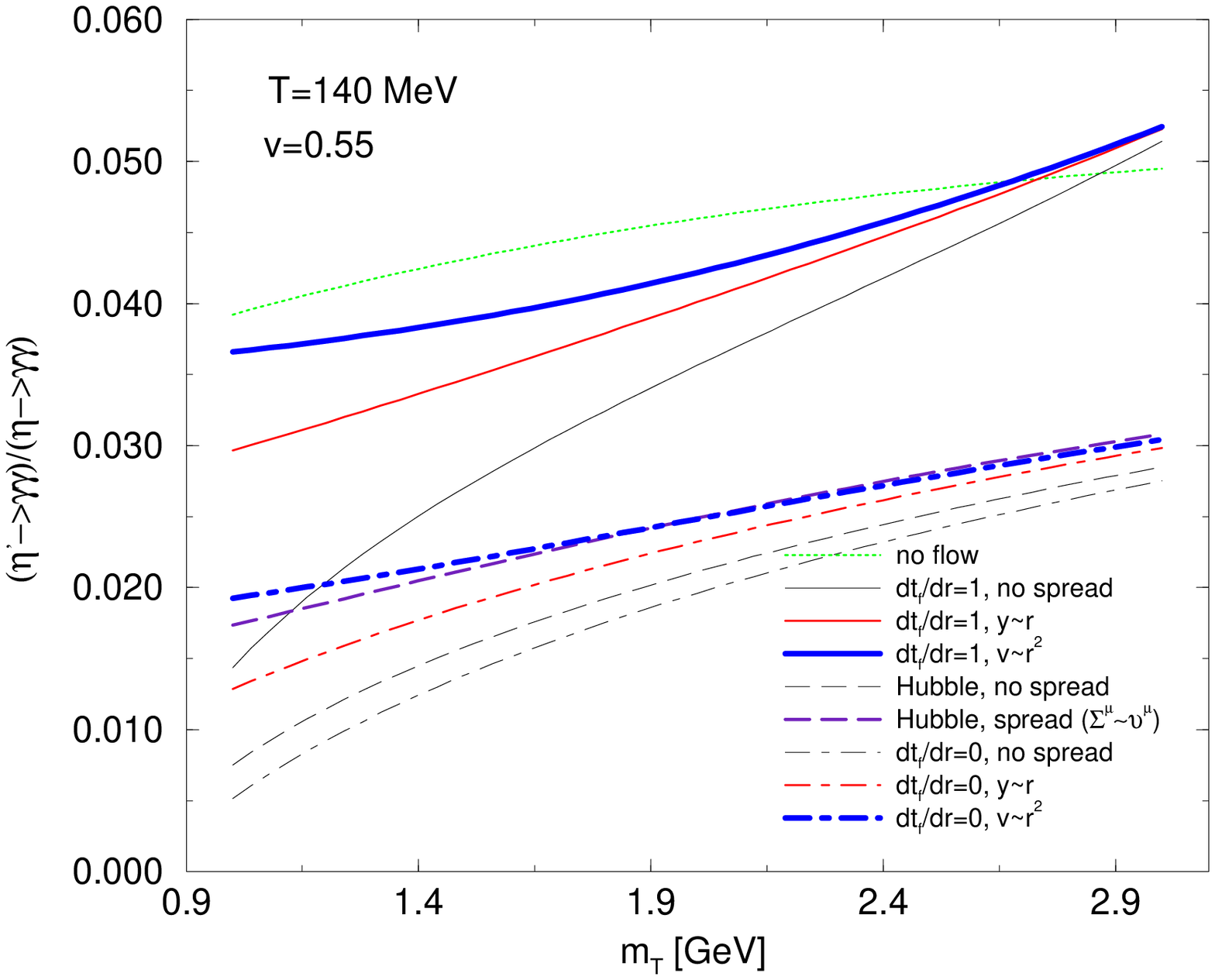}
}
\resizebox*{!}{0.32\textheight}{
\includegraphics{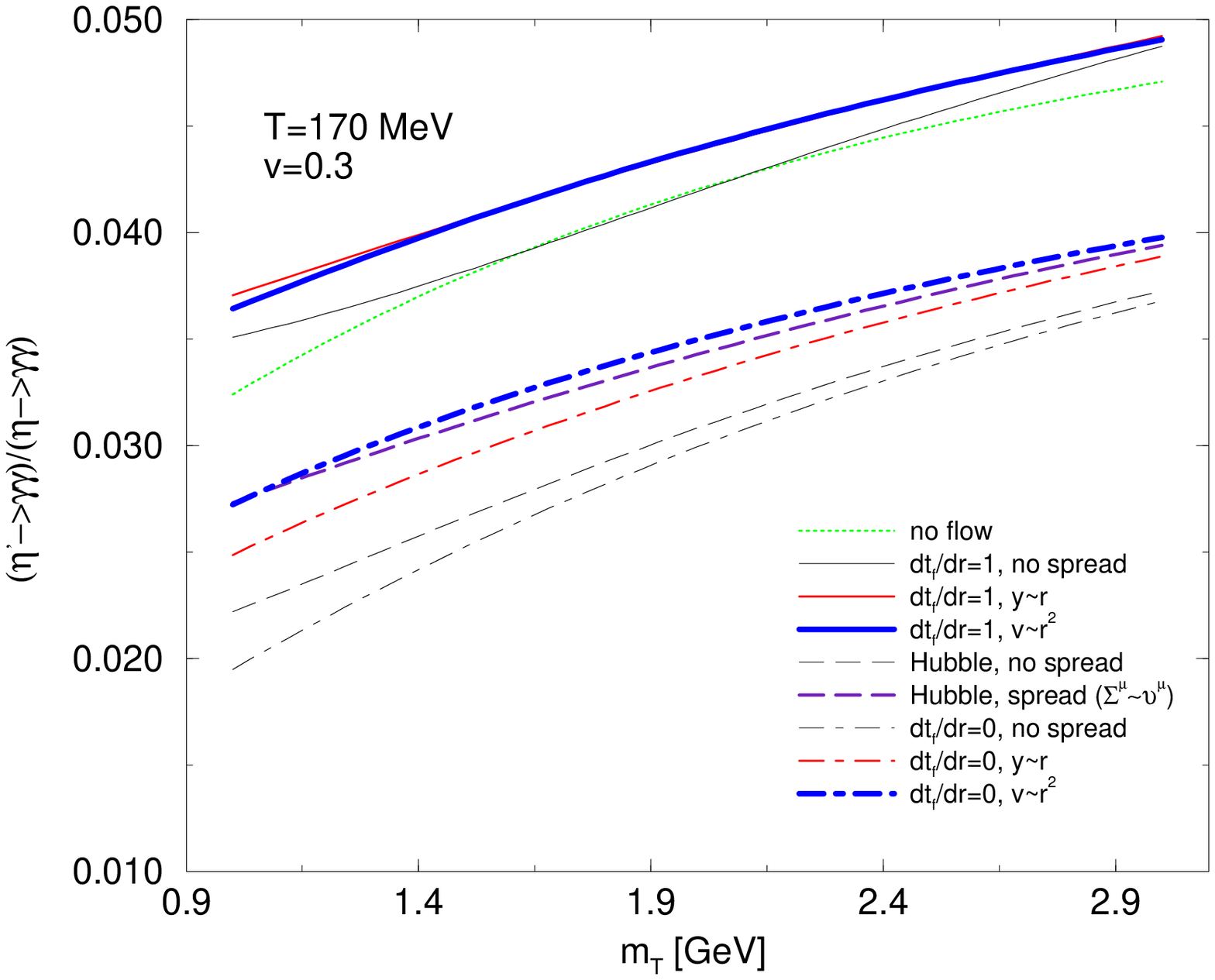}
}}
\end{center}
\caption{(Color online) $(K^*+\overline{K}^*)/($all $K_S)$, $\Sigma^*(1385)/($all $\Lambda)$ 
and $\eta'/($all $\eta)$ ratios, including feed down from resonances.
\label{diagfeed} }
\end{figure*}

Figure ~\ref{diagfeed} shows the ratios, including feed down
of resonances, for the same particles and statistical hadronization 
conditions  as were studied in Fig.~\ref{diagres}.
In the $\Sigma^*/($all $\Lambda)$ case we omitted the feed down from $\Xi$ to 
$\Lambda$ which is usually corrected for (if this is not done
the ratio  $\Sigma^*/($all $\Lambda)$ would depend strongly on the chemical potentials).
We did allow for the $\phi \rightarrow K_S K_L$ feed down, since it is a 
strong decay that cannot so easily be corrected for. We note that 
the feed down from particles with a different chemical composition
cannot always be corrected for, and thus some resonances ratios will also acquire a 
(mild) dependence on the chemical potentials.
This is even  true for ratios such as $\eta'/($all $\eta)$, given 
different $s \overline{s}$ content,
in this paper, these type chemical corrections were set equal to unity.

\begin{figure*}[tb]
\begin{center}
\centerline{\resizebox*{!}{0.3\textheight}{
\includegraphics{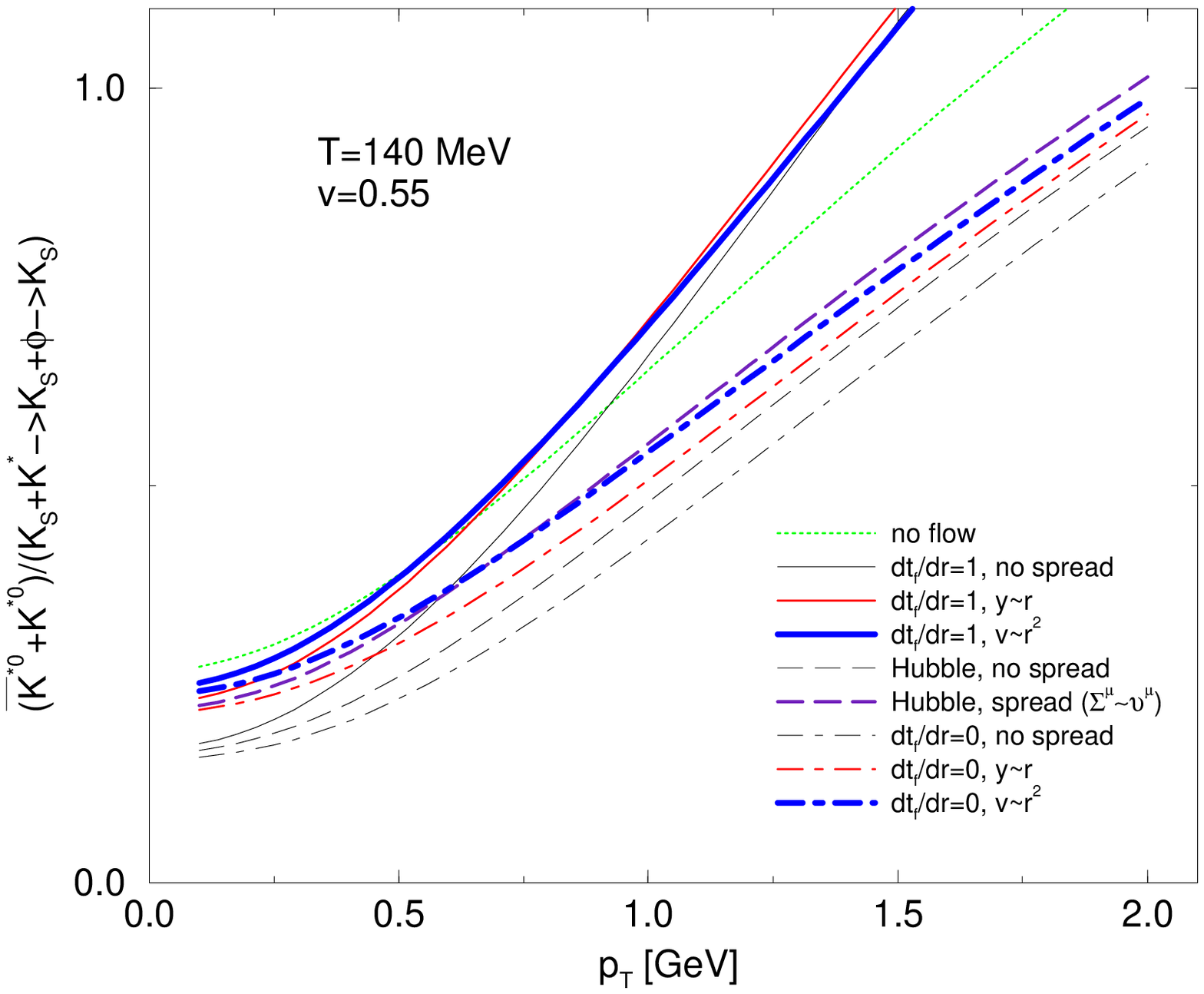}
}
\resizebox*{!}{0.3\textheight}{
\includegraphics{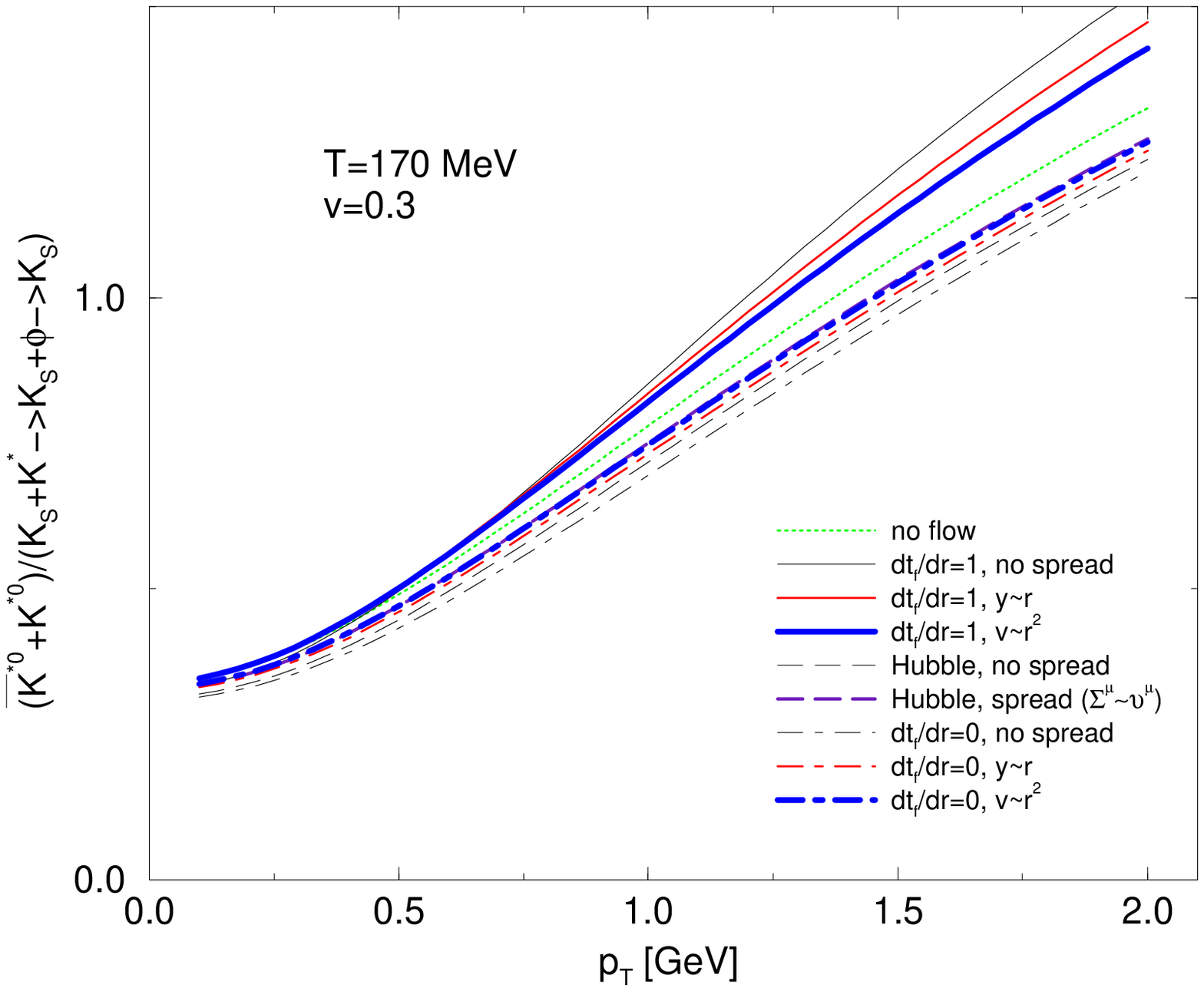}
}}
\centerline{\resizebox*{!}{0.3\textheight}{
\includegraphics{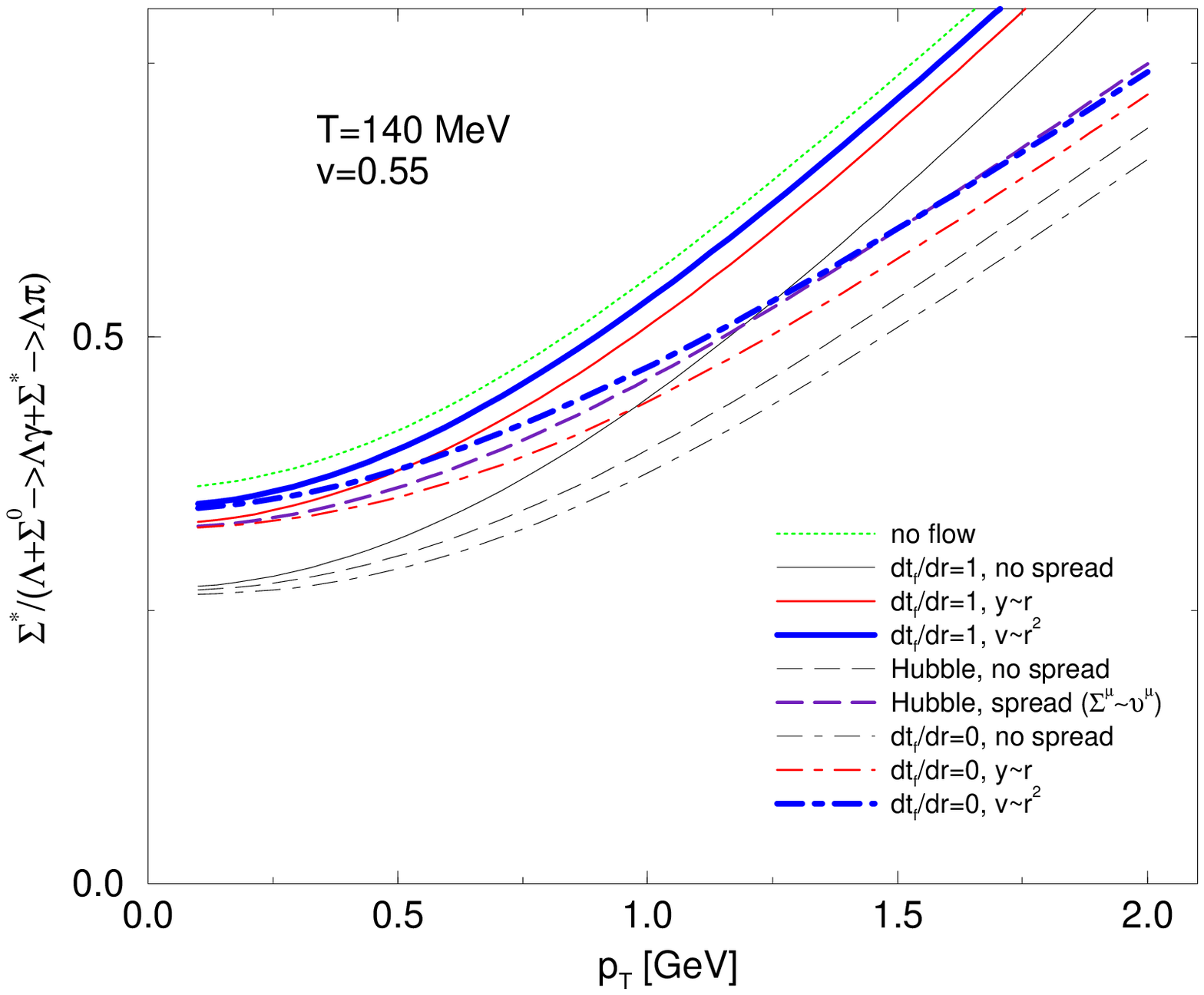}
}
\resizebox*{!}{0.3\textheight}{
\includegraphics{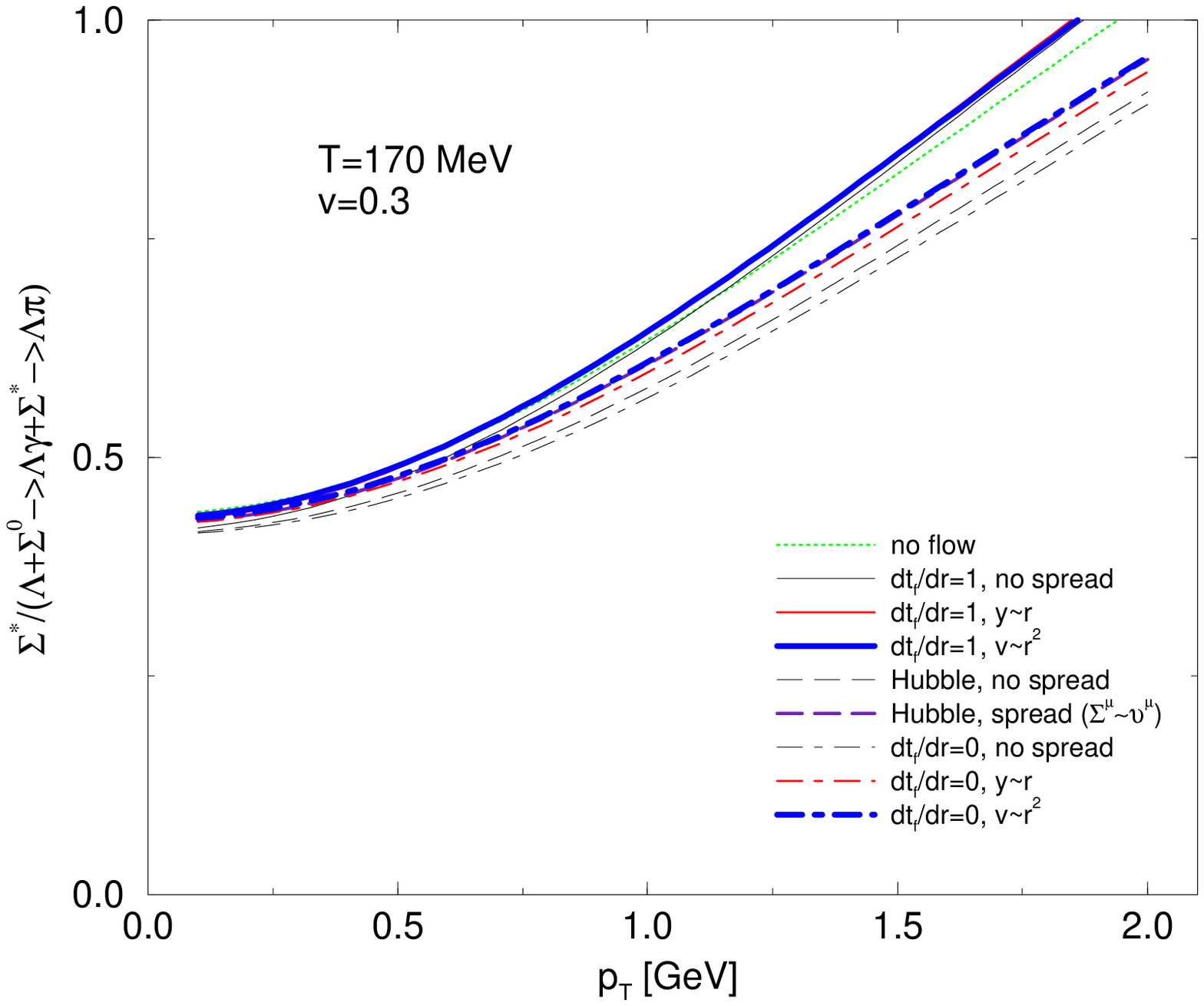}
}}
\centerline{\resizebox*{!}{0.28\textheight}{
\includegraphics{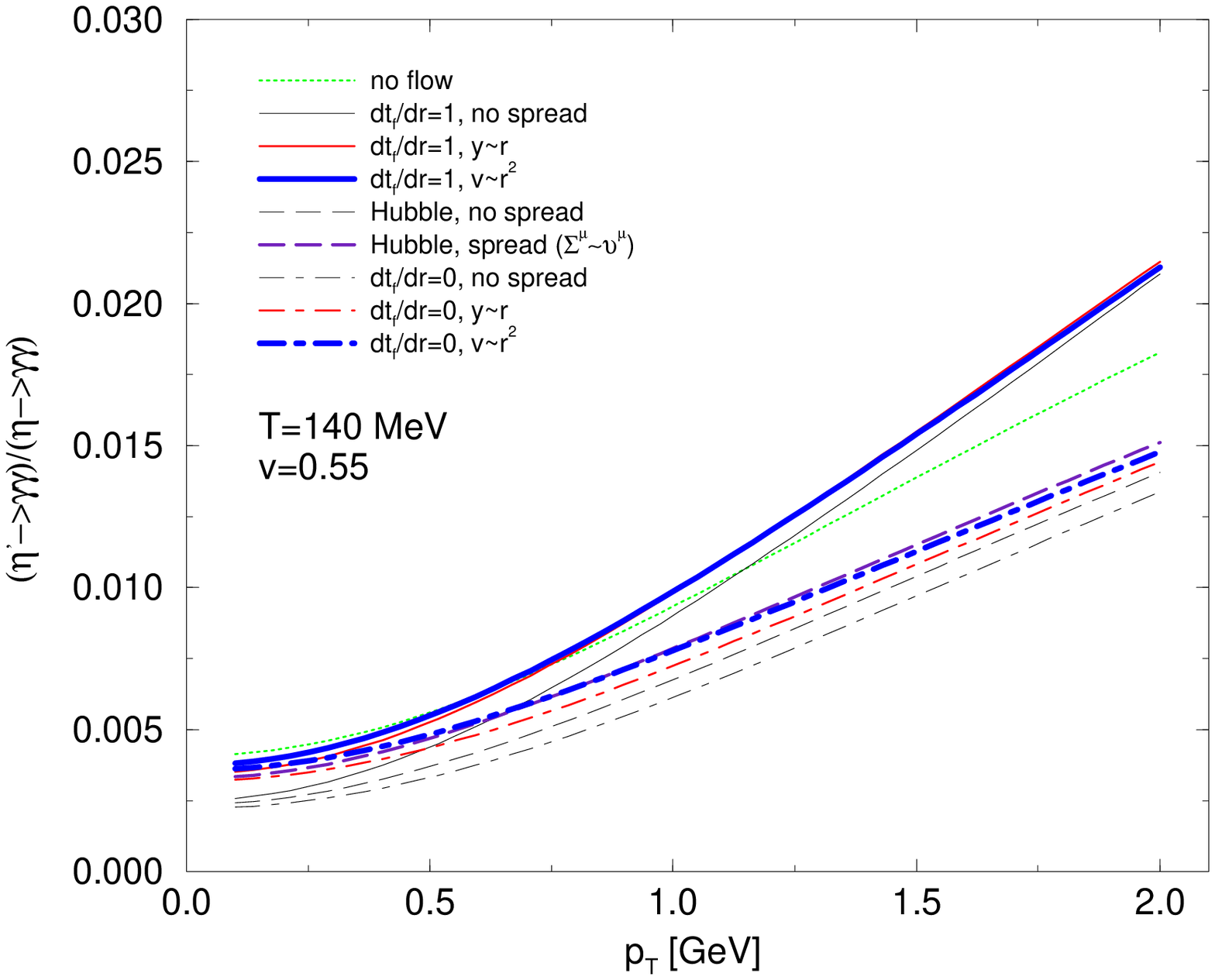}
}
\resizebox*{!}{0.28\textheight}{
\includegraphics{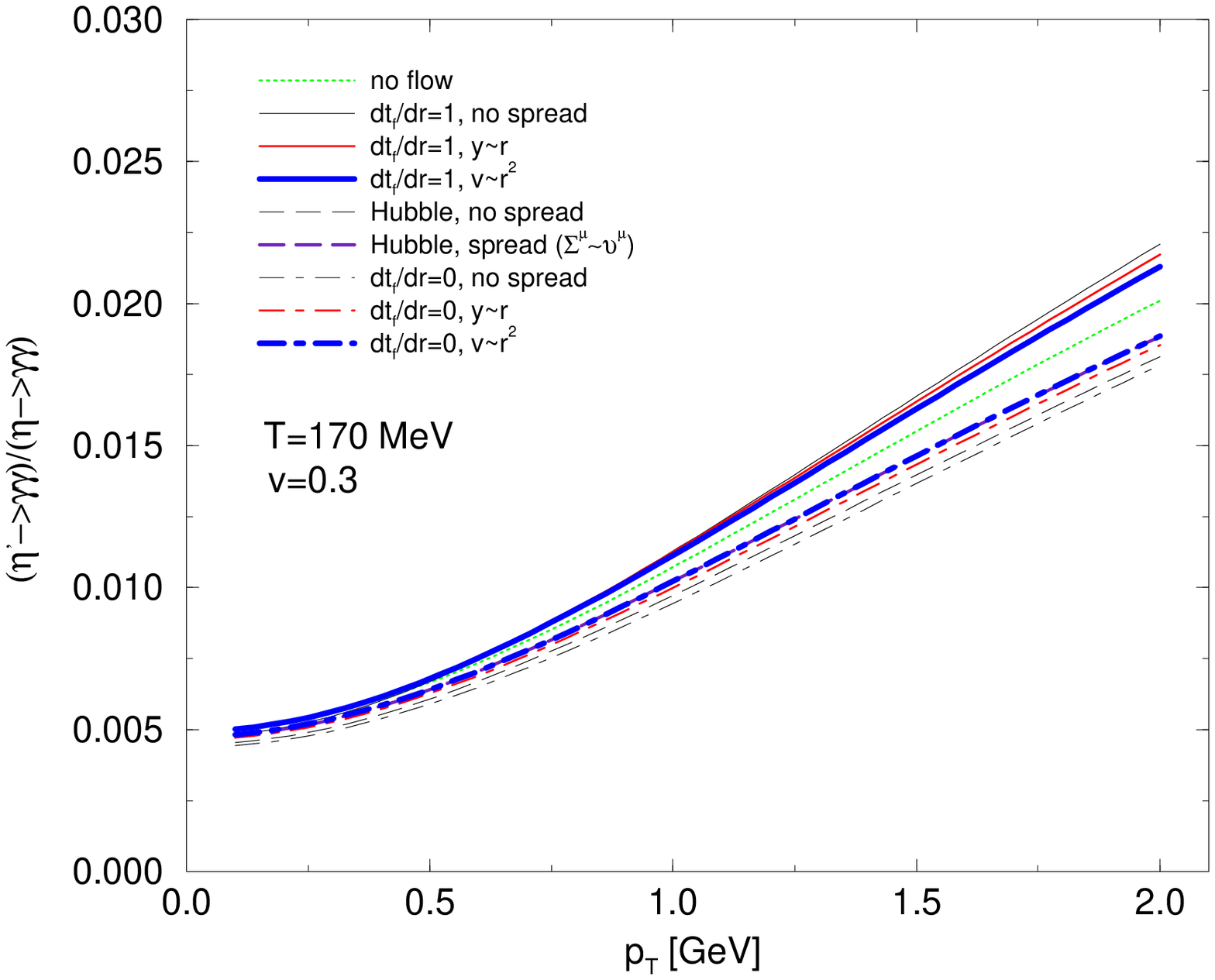}
}}
\end{center}
\caption{(Color online) $p_\bot$ dependence of $(K^*+\overline{K}^*)/($all $K_S)$, $\Sigma^*(1385)/($all $\Lambda)$ 
and $\eta'/($all $\eta)$ ratios, including feed down from resonances.
\label{diagfeedpt} }
\end{figure*}

To further study the sensitivity of resonance-particle $m_\bot$-ratio 
to freeze-out dynamics, we also present the (feed down corrected)  case 
as a function of $p_\bot$ rather than $m_\bot$ in Fig.~\ref{diagfeedpt}.
Unsurprisingly, we see grossly different behaviors, with many of the results
coalescing. This of course  is an expression of the fact that 
$Y^*$ and $Y$  have dramatically different $p_\bot$ at the same $m_\bot$ and vice versa.
We believe  that the $m_\bot$ ratio will in general be more sensitive to freeze-out 
dynamics, since its dependence on $m_\bot$ is dominantly  due to freeze-out geometry 
and dynamics. 
However, the $p_\bot$ dependence seen in Fig.~\ref{diagfeedpt} provides
 an important self-consistency check for our previous results. We have found that
the $m_\bot$ ratios are often greater than unity even though  
there must be more ground state
 particles than resonances. Now it can be  seen in the  
$p_\bot$ ratio,  that this requirement is satisfied. 

\section{Discussion}
In general the the $m_T$ and $p_T$ dependence of the ratios in Fig 3 and, respectively, Fig
4 depends on freeze-out geometry and dynamics. 
Changes in temperature and flow velocity  alter the shape. 
The introduction of a steeper flow profile will further  raise 
all of the considered ratios,
since a considerable fraction of particles will be produced in regions that do
not flow as much.
The effect of freeze-out dynamics will generally go in the same direction
as freeze-out approaches the explosive limit ($dt_f/dr\rightarrow1$).
However, both the magnitude and the qualitative features of the two 
effects (flow and freeze-out velocity) 
will be considerably different. Especially, when more than one ratio is 
measured, it would appear that we will be able to 
determine the freeze-out condition. 
This is in contrast to the $m_\bot$ distributions in Fig.~\ref{mcplots}, where 
the effects discussed in this paper result in linear corrections, which tend to
compete, making the task of extracting the freeze-out dynamics much more
ambiguous. Thus, there is considerable 
potential of resonance-particle $m_\bot$-ratios as a freeze-out 
probe.
 
The presence of a long living hadronic gas rescattering phase
can distort our freeze-out probe.
In particular, the apparent  $Y^*/Y$ ratio will be altered due to the depletion of
the detectable resonances through the rescattering of their decay products.
It's dependence on $m_\bot$  will  be affected in a non-trivial
way,  since faster (higher
$p_\bot$) resonances will have a greater chance to escape the fireball
without decaying, thus avoiding the rescattering phase altogether.
Regeneration of resonances in hadron scattering may add another  
$m_\bot$ dependence which is different for the $\Sigma^*/\Lambda$ and
the $K^*/K$ ratios~\cite{bleicher}. Other signals of the existence of such an interacting
hadron gas phase have been considered~\cite{sqm2001,resonances}.
 Fortunately, there is no evidence that
a rescattering phase  plays a great role in particle distributions.
Even so, it would seem that the ``safest''  probes for freeze-out are
 the particles  and resonances most unlikely to rescatter.

For this reason we have included the $\eta'/\eta$ ratio in our considerations. 
$\eta\rightarrow \gamma \gamma$ and $\eta'\rightarrow \gamma \gamma$ have very 
different branching ratios,
but  have the same degeneracies and similar but rather  small partial  widths.
The electromagnetic decay mode is practically  insensitive to posthadronization
dynamics. Regeneration effects are suppressed since the hadronic two body decay channel
is suppressed. 
All these features make these particles interesting  probes, allowing 
for the analysis considered here.
$\eta,\eta'$ mesons have been measured  at SPS energies 
in the $\gamma\gamma$ decay channel~\cite{eta1,eta2}, and detectors such as PHENIX
are capable of reconstructing the same decays at RHIC.  

While a long rescattering phase would affect the $\Sigma^*$ distribution, 
the effect would be very easy to detect experimentally:
$95 \%$ of $\Sigma^*$ decay through the $p$-wave $\Sigma^* \rightarrow \Lambda \pi$ channel.    However, regenerating $\Sigma^*$ s in a gas of $\Lambda$ s and
$\pi$ s is considerably more difficoult, since $\Lambda \pi$ scattering
will be dominated by the s-wave
$\Lambda \pi \rightarrow \Sigma^{\pm}$.   
This situation will not occur for
$K^* \leftrightarrow K \pi$, since both decay and regeneration happen 
through the same process, leading to a very fast reequilibration
time \cite{bleicher}.  
Since both $\Sigma^* /\Lambda$ and $K^*/K$ ratios have been calculated
within the thermal model \cite{sqm2001} (neglecting rescattering), a
strongly depleted $\Sigma^*/\Lambda$ ratio (compared with $K^*/K$ )
suggests that a statistical freeze-out description,
such as that given in this paper, is incomplete, and an interacting
hadron gas phase is also necessary.

In summary, we have presented an overview of the different statistical 
freeze-out models used to fit heavy ion data. We have shown how the 
freeze-out geometry and freeze-out dynamics influences the hadron spectra.
Our primary result is the finding that the 
 $m_\bot$ dependence of the resonance-particle  ratios is a probe 
of freeze-out. We have presented these ratios for three particle species and 
two freeze-out conditions and have considered how our results could 
be altered by posthadronization phenomena. 

\acknowledgments
Supported  by a grant from the U.S. Department of
Energy,  DE-FG03-95ER40937\,. 
We thank Patricia Fachini (BNL) and Zhangbu Xu (Yale), from the
STAR collaboration, as well as Marcus Bleicher (ITP-Frankfurt)
 for fruitful discussions.

\end{document}